# The Anthropocene by the Numbers: A Quantitative Snapshot of Humanity's Influence on the Planet

Griffin Chure[1,a], Rachel A. Banks[2], Avi I. Flamholz[2], Nicholas S. Sarai[3], Mason Kamb[4], Ignacio Lopez-Gomez[5], Yinon M. Bar-On[6], Ron Milo[6], Rob Phillips[2,4,7,*]

[1]Department of Applied Physics and Materials Science, California Institute of Technology, Pasadena, CA, USA
[2]Division of Biology and Biological Engineering, California Institute of Technology, Pasadena, CA, USA
[3]Division of Chemistry and Chemical Engineering, California Institute of Technology, Pasadena, CA, USA
[4]Chan-Zuckerberg BioHub, San Francisco, CA, USA
[5]Department of Environmental Science and Engineering, California Institute of Technology, Pasadena, CA, USA
[6]Department of Plant and Environmental Sciences, Weizmann Institute of Science, Rehovot, Israel
[7]Department of Physics, California Institute of Technology, Pasadena, CA, USA

[a]Current Address: Department of Biology, Stanford University, Stanford, CA, USA

## Abstract

The presence and action of humans on Earth has exerted a strong influence on the evolution of the planet over the past ≈ 10,000 years, the consequences of which are now becoming broadly evident. Despite a deluge of tightly-focused and necessarily technical studies exploring each facet of "human impacts" on the planet, their integration into a complete picture of the human-Earth system lags far behind. Here, we quantify twelve dimensionless ratios which put the magnitude of human impacts in context, comparing the magnitude of anthropogenic processes to their natural analogues. These ratios capture the extent to which humans alter the terrestrial surface, hydrosphere, biosphere, atmosphere, and biogeochemistry of Earth. In almost all twelve cases, the impact of human processes rivals or exceeds their natural counterparts. The values and corresponding uncertainties for these impacts at global and regional resolution are drawn from the primary scientific literature, governmental and international databases, and industry reports. We present this synthesis of the current "state of affairs" as a graphical snapshot designed to be used as a reference. Furthermore, we establish a searchable database termed the Human Impacts Database (www.anthroponumbers.org) which houses all quantities reported here and many others with extensive curation and annotation. While necessarily incomplete, this work collates and contextualizes a set of essential numbers summarizing the broad impacts of human activities on Earth's atmosphere, land, water, and biota.

## Introduction

One of the most important scientific developments in modern history is the realization that the evolution of the Earth is deeply intertwined with the evolution of life. Perhaps the most famous example of this intimate relationship is the chemical transformation of Earths' atmosphere following the emergence of photosynthesis, an event so important to Earth's history that it has been colloquially termed the "Oxygen Holocaust" due to the massive extinction event that followed[1–3]. Over the past ≈ 10,000 years, humans have become a similarly influential force of nature, directly influencing the rise and fall of ecosystems[4–12], the temperature and volume of the oceans[13–19], the composition of terrestrial biomass[20], the planetary albedo and ice cover[21–28], and the chemistry of the atmosphere[29–34] to name just a few of many such examples. The breadth of human impacts on the planet is so diverse that it penetrates nearly every scientific discipline.

This penetration has resulted in a deluge of data, allowing us to describe the many facets of human impacts in quantitative terms often with remarkable precision and high resolution. However, these works are typically highly technical and tightly-focused, meaning that they paint a fragmented picture of the sweeping global changes wrought by human activities. Even seemingly simple questions such as "how much land or water do humans use?" can be difficult to answer when searching the scientific literature yields an array of complicated analyses with inconsistent definitions, methods, and assumptions, each reporting their findings in different units. This problem can lead to the perception that there is a disagreement on the facts when in reality there is broad scientific consensus.

We present a synthesis of these data, drawing heavily from scientific studies as well as industrial and governmental databases to assemble a broad, quantitative view of human action on Earth in a set of consistent and intuitive units. Paul Crutzen, who famously coined the term Anthropocene[35] in 2002, captured the essence of human impacts by pointing out that humans have transformed between 30-50% of the Earth's land area and increased the atmospheric concentration of $CO_2$ by 30% over the years 1800-2000. In this same spirit, we formulate and calculate an array of dimensionless quantities that compare the magnitude of the human impact (such as the biomass of livestock) to the natural analogue (such as the biomass of wild terrestrial animals). To make the data on this broad array of impacts transparent and their uncertainties clear, we have established a searchable online database which acts as a repository for the values presented here as well as many others. This manually curated and continuously updated database, termed the Human Impacts Database ([www.anthroponumbers.org](www.anthroponumbers.org)), contains a variety of quantities, each with a unique 5-digit identifier termed an HuID, which we reference throughout this work.

We consider the act of assembling the facts and making them easily accessible a prerequisite to discussing the actions that should be taken to manage the effects of human impacts. However, our work has led to a number of insights not obvious when the impacts are considered in isolation. First, as will be seen throughout the paper, though human activities impact a wide array of natural processes, these impacts are intimately intertwined in much the same way that the biosphere is intertwined with the atmosphere, geosphere and hydrosphere. Over and over, we find that the scale of human impacts is driven by a small number of crucial factors: the size of the human population and our demand for power, water and food. Second, by comparing the scale of human impacts (such as total biomass of livestock) to their corresponding natural counterparts (total biomass of wild animals) we see that human actions rival or exceed their natural analogues in nearly all cases.

## Results
### *Global Magnitudes*

In Figure 1, we present a gallery of critical human impacts at the global scale, categorized into five classes: land, water, flora and fauna, atmospheric and biogeochemical cycles, and energy, as indicated by the corresponding color of their banners. Though the impacts considered are necessarily incomplete, these values encompass many critically important numbers, such as the volume of liquid water resulting from ice melt (Figure 1 B), the extent of urban and agricultural land use (Figure 1 H), global power consumption (Figure 1 N), and the heat uptake and subsequent warming of the upper ocean (Figure 1 S). We direct the reader to the supplemental information for a detailed description and full referencing of each number reported in Figure 1.

Exploring these numbers reveals surprising quantities and relationships. For example, agriculture is a major contributor to many of these impacts, dominating both global land use (Figure 1 H, HuID: 29582) and human water use (Figure 1 L; HuIDs: 84545, 43593, 95345), as well as accounting for a third of global tree cover area loss (Figure 1 O; HuID: 24388). In addition, a vast amount of nitrogen is synthetically fixed through the Haber-Bosch process to produce fertilizer (Figure 1 F; HuID: 60580, 61614), which is a major source of $N_2O$ emissions (Figure 1 K; HuID: 44575). Our collective agricultural practices also make us the stewards of a standing population of around 30 billion livestock (Figure 1 E; 15765). Along with rice paddies, ruminant livestock like cattle produce a majority of anthropogenic methane emissions (Figure 1 K; HuID: 96837, 30725). In contrast, urban land area accounts for a nearly negligible fraction of land use (Figure 1 H; HuID: 41339, 39341), and urbanization accounts for only ≈ 1% of global tree cover area loss (Figure 1 O; HuID: 19429). This is not to say, however, that urban centers are negligible in their global impacts. Urban areas now house more than half of the global human population (Figure 1 J; HuID: 93995) and the construction of urban structures, as well as the roads, tunnels, dams, and factories supporting them, is the dominant cause of earth-moving operations on an annual basis (Figure 1 W; HuID: 59640).

Collectively, the ≈ 8 billion humans on Earth (Figure 1 J; HuID: 85255) consume nearly 20 TW of power (Figure 1 N; HuID: 31373, 85317), with around 80% coming from the combustion of fossil fuels (Figure 1 P;  HuID: 29470, 29109). This results in a tremendous mass of carbon dioxide emitted annually (Figure 1 K; HuID: 24789, 54608, 98043), which, along with other greenhouse gases, has increased the global surface temperature via the greenhouse effect by more than  1° C relative to the average temperature between 1850 and 1900 (Figure 1 A; HuID: 79598, 76539, 12147). A sizable portion of these emissions are absorbed by the oceans, leading to a steady increase in ocean acidity (Figure 1 G; HuID: 90472, 19394), threatening many marine ecosystems[36–41]. Furthermore, increasing average global temperatures contributes to sea level rise not only in the form of added water from ice melt and discharge from ice sheets (Figure 1 B and M; HuID: 32459, 95978, 93137, 81373, 70818) but also through the thermal expansion of water, which accounts for ≈ 30% of the observed global sea level rise (Figure 1 M; HuID: 97688).  These are just a few ways in which one can traverse the impacts illustrated in Figure 1, revealing the remarkable extent to which human impacts are interconnected. In the supplemental information, we outline several other paths one could take through Figure 1 and  encourage readers to explore this figure in a similar manner.

While Figure 1 presents the magnitude of human impacts at a global scale, it is important to recognize that both the origins and repercussions of human activities are highly variable across the globe. In the supplemental information and supplemental Figure S1, we present coarse-grained regional breakdowns of many of the  numbers from Figure 1 for which regional distributions could be determined.

## *Dimensionless Representations*

To quantitatively describe the Anthropocene, it is useful to compare the magnitude of human impacts to  their natural analogues in a spirit similar to that initially presented nearly 20 years ago by Paul Crutzen[35]. In Figure 2, we outline a dozen "dimensionless ratios" which compare contemporary human impacts to natural processes closely related in scope or in value. We begin with one of the central questions pertaining to the anthropocene -- the extent of human land use. *The Terra Number,* diagrammed in Figure 2 A compares human terrestrial land use to total global terrestrial land area. This comparison reveals that humans have transformed 30% of terrestrial land for agricultural (≈ 5 × $10^7$ $km^2$, HuID: 29582) and urban developments  (≈ 7 × $10^5$ $km^2$; HuID: 41339, 39341). In many cases, utilizing this land requires clearing it of its natural biota, such as trees and shrubs, in order to grow food or build structures.

*The Deforestation Number***,** shown in Figure 2 B, puts the extent of such land clearing in context.  Through a combination of permanent deforestation (e.g. clear-cutting forest where there is no regrowth of natural tree cover) and temporary tree cover loss (such as tree cover lost through managed forestry and agriculture, where there is eventual regrowth), humans intentionally remove tree cover. Through these intentional and managed means, humans  clear an area of ≈ 1.8 × $10^5$ $km^2$ annually (HuID:  96098 , 24388, 38352, 19429), approximately twice as large as the area cleared through wildfires  (≈ 7 × $10^4$ $km^2$; HuID: 92221), some of which are human-caused (a subset of which are controlled burns). Agricultural activities alone lead to the clearing of an area comparable to that of wildfire. This cleared land is used not only to grow food but rear an impressive number of animals as livestock.

The enormity of the standing terrestrial livestock population is contextualized by *The Barnyard Number*, illustrated in Figure 2 C. The total  biomass of global terrestrial livestock is   ≈ 2 × $10^{12}$ kg and currently outweighs all terrestrial wild mammals and wild birds (≈ 7 × $10^{10}$ kg)[20]  by approximately 30 fold. On a mass basis, therefore, it is more realistic to picture dominant land animals as cows and chickens rather than elephants and zebras. Together, the ≈ 30 billion terrestrial livestock animals (HuID: 43599) who make up this reservoir of biomass are dependent on humans to grow their food, culminating in an enormous demand for nitrogenous fertilizer and water.

The production of synthetic fertilizer requires chemical fixation of nitrogen to make reactive species like ammonia ($NH_3$), which plants can utilize, from the inert dinitrogen gas ($N_2$) that makes up nearly 80% of our atmosphere. Natural $NH_3$ production is catalyzed by various species of bacteria that often form symbioses with plants[42], however natural formation of $NH_3$ is not sufficient to sustain a human population beyond ≈ 3 billion individuals[43]. To meet this demand for $NH_3$, humans synthesize ≈ $1.5 \times 10^{11}$ kg of reactive nitrogen annually via the energy-intensive Haber-Bosch process (HuID: 60580, 61614), a mass equal to that of all biological nitrogen fixation on Earth. The observation that humanity now matches nature in terms of production of reactive nitrogen is reflected in *The Nitrogen Number* calculated in Figure 2 (D). Since humanity now applies ≈ $10^{11}$ kg of nitrogenous fertilizer globally each year, it is crucial to ensure that the fertilized crops also receive enough water.

*The Water Number (Figure 2 E)* reflects the total amount of water withdrawn by humans, which is dominated by irrigation of cropland (≈ $1.5 \times 10^{12}$ m$^3$ / yr; HuID: 43593), and closely followed by industrial use (≈ $6 \times 10^{11}$ m$^3$ / yr; HuID: 27142), namely in the form of electricity production. Together, these two cases account for ≈ 95% of total human water use per year (HuID: 84545, 43593, 95345, 27142, 27342, 68004). Total human water withdrawals are about 5% of global river discharge volume, the major source of renewable freshwater[44]. While this is a small fraction of available freshwater, freshwater is highly variable across the globe and about a third of the human population lives in water stressed areas, where water withdrawal is greater than 40% of available freshwater[44].

*The River Number*, shown in Figure 2 F, summarizes the extent to which humans have shaped the flow of freshwater, harnessing it for irrigation and hydropower, and changing the flow to prevent floods. Currently, the total volume of rivers under human control (≈ $6 \times 10^{11}$ m$^3$; HuID: 61661) , by dams and myriad smaller barriers, equals the total volume of free flowing rivers (≈ $6 \times 10^{11}$ m$^3$; HuID: 55718). This massive repurposing of water flow has a huge impact on not only the water cycle but also the movement of aquatic organisms and sediment. Previous to human action, rivers were the major force moving sediment, but humans have now taken over that role.

The vast amount of sediment moved by humans is encompassed by *The Earth Mover Number* (Figure 2 G). Currently, humans move at least $2.5 \times 10^{14}$ kg / year (HuID: 19415, 59640, 72899) over 15 times the amount of sediment that is moved by rivers (≈ $1.3 \times 10^{13}$ kg / year, HuID: 51481). Urbanization alone accounts for over half of this, driving the movement of over $1.4 \times 10^{14}$ kg of sediment a year (HuID: 59640). Waste and overburden from coal mining and erosion due to agriculture make up the remaining ≈ $0.9 \times 10^{14}$ kg / year (HuID: 19415, 41496, 72899). In addition to a massive amount of earth moved, urbanization also requires an astounding amount of human-made materials such as steel and concrete.

Our calculation of *The Anthropomass Number* (Figure 2 H) shows that the mass of human-made materials now equals the dry weight of all living matter on the planet[45]. Material of human origin, termed *anthropomass* is dominated by construction materials, especially concrete (≈ $3 \times 10^{13}$ kg / yr; HuID: 16995, 25488, 81346) and other aggregates (asphalt, sand, gravel, and bricks), with steel coming next (≈ $1.9 \times 10^{12}$ kg / yr; HuID: 44894, 51453, 85891). In addition to the raw mineral resources, producing these materials requires an enormous amount of energy.

$CO_2$ is the most famous greenhouse gas and also the one that human activities produce in the largest quantity. *The $CO_2$ Number* (Figure 2 J) quantifies the amount of $CO_2$ produced by human processes relative to the natural sinks, namely absorption by Earth's oceans and photosynthesis by plants and algae[29]. About 85% of anthropogenic $CO_2$ emissions (≈ $4 \times 10^{13}$ kg / yr; HuID: 60670, 24789, 54608) are due to burning fossil fuels like coal, oil, and natural gas, with the remainder due to land-use changes like the removal of forests for agriculture. Annual anthropogenic $CO_2$ emissions are roughly double the amount removed by natural sinks ( ≈ $2 \times 10^{13}$ kg / yr; HuID: 52670). The remainder remains in the atmosphere, growing in concentration year after year. The excess $CO_2$ in the atmosphere contributes to warming the planet, while $CO_2$ absorbed by oceans changes the pH, increasing its acidity and disrupting oceanic ecosystems[21]. $CO_2$ is not, however, the only greenhouse gas humans produce in significant quantities.

Methane is an extremely potent greenhouse gas, having a global warming potential about 25 times that of $CO_2$ over 100 years[46]. Anthropogenic $CH_4$ emissions are mainly due to cattle, rice paddies, and burning fossil fuels[34] . As shown by *The $CH_4$ Number* (Figure 2 K) natural (≈ 3 × $10^{11}$ kg / yr; HuID: 56405) and anthropogenic methane emissions (≈ 4 × $10^{11}$ kg / yr; HuID: 96837) are currently roughly equal, indicating that humanity matches nature in the production of this critical gas.

The mining and burning of fossil fuels for electrical and mechanical power is a major driver of both $CO_2$ and $CH_4$ emissions. Despite using an enormous 18 TW of power globally (HuID: 31373, 85317), humanity only consumes a miniscule fraction of the power incident on the planet from the sun. This fraction is calculated as *The Solar Number* (Figure 2 I), which shows that current human power consumption is roughly 0.01% of annual incident solar energy. Today, most of our power derives from burning fossil fuels and only about 1% is solar. Solar power is therefore an enormous and mostly-untapped resource.

The 12 dimensionless numbers presented in Figure 2 quantify diverse yet specific ways in which humans are influencing the evolution of the Earth. The processes described by these 12 ratios are distinct, but interrelated, sharing many common themes. Intensification of agriculture leads us to convert forest to cropland, divert rivers and burn fossil fuels to power the synthesis of nitrogenous fertilizer. Building roads, tunnels, dams, homes and offices leads to massive earth-moving operations that similarly demand water, power, and land area. These myriad interconnected human activities have complex effects on natural ecosystems, many of which are smaller and more fragmented than they were 200 years ago. Our final dimensionless ratio, *The Extinction Number* (Figure 2 L), attempts to understand the scale of these effects on ecosystems. Over the past 500 years, at least ten times more animal species have gone extinct (≈ 760 species; HuID: 44641) than would be expected given the most conservative estimate for the background extinction rate[8,47]. The causes of this increased extinction rate are varied and remain poorly understood, and this quantity is necessarily a lower bound as only a small fraction of species have been studied. It is very likely that extinction is far more prevalent than we know, especially among arthropods and marine biota more generally.

## Discussion

In this work we canvassed the scientific literature as well as governmental and international reports to assemble a broad, quantitative picture of how human activities have impacted Earth's atmosphere, oceans, rives, lands, biota and geology. We assembled these data into a comprehensive snapshot, released alongside this writing as a standalone graphical document (Supplemental Material 1) , with all underlying data, associated uncertainties and referencing housed in the Human Impacts Database. As illustrated by the dimensionless ratios presented in Figure 2, the scale of these impacts is not small. Rather, in nearly all cases, human activities impact the planet to a degree rivalling or even exceeding counterpart natural processes. Perhaps even more so than any other ratio we present, *The Anthropomass Number* (Figure 2 H) conveys this point by showing that the total mass of human-made "stuff" roughly equals the total mass of all living matter on Earth[45].

One insight that emerges from considering these diverse human activities together is that they are deeply intertwined and driven by a small number of pivotal factors: the size of the human population, the composition of our diets, and our demand for materials and energy to build and power our increasingly complex and mechanized societies. Understanding the scale of human agriculture, water and power usage provides a framework for understanding nearly the entirety of the numerical gallery presented in Figure 1. Indeed, agriculture alone is the dominant cause of human land and water use and a major driver of deforestation as well as methane and nitrous oxide emissions.

It is common in this setting to argue that the bewildering breadth and scale of human impacts should motivate some specific remediation at the global or local scale. We prefer a more modest "just the facts" approach. The numbers presented here show that human activities affect our planet to a large degree in many different and incommensurate ways, but they do not provide a roadmap for the future. Rather, we contend that any plans for the future should be made in the light of a comprehensive and quantitative understanding of the interconnected ways in which human activities impact the Earth system globally (Figures 1 and 2) and locally (Figure S1). Achieving such an understanding will require synthesis of broad literature across many disciplines, work that we have only just begun. While the quantities we have chosen to explore are certainly not exhaustive, they represent some of the key axes which frequently drive scientific and public discourse and shape policy across the globe.

The quantitative picture presented here summarizes the "state of affairs" as of 2020. Earth is the only habitable planet we know of, so it is crucial to understand how we got here and where we are going. That is, how have human impacts changed over time? How are they expected to change in the future? For every aspect of human entanglement with the Earth environment – from water use to land use, greenhouse gas emissions, mining of precious minerals, and so on – there are excellent studies measuring impacts and predicting their future trajectories. The time has come to integrate these disparate works and synthesize a complete picture of the human-Earth system, one that helps humanity coexist stably with the only planet we have.

## Materials & Methods

All values reported in this work come from myriad sources in the scientific, industrial, governmental, and organizational reports and articles. Every value reported in Figure 1 as well as Supplemental Figure S1 are extensively documented in the supplemental information. In brief, each value reported was manually curated with one or more of the authors closely reading the original report/article/document, downloading (or in some cases requesting) the original datasets, and annotating and cleaning the data to follow the principles of "tidy data" formatting[48]. For each reported quantity, we identified the method of determination (e.g. direct measurement, statistical inference, or aggregated estimate) and provided an assessment of uncertainty. In cases where the uncertainty was undetermined or not reported, we sought additional sources to provide a range of reasonable values. Thus, as is reported in the supplemental information, we use equality symbols (=) to indicate values with an estimate of the uncertainty or values that are tightly constrained in range, and approximation symbols (≈) to indicate estimates likely accurate to within a factor of a few, and an inequality symbol (>) to indicate a lower-bound estimate for the value in question. Every dataset and source we considered in this work has been stored in a GitHub Repository (DOI: 10.5281/zenodo.4453277; https://github.io/rpgroup-pboc/human_impacts) with extensive documentation. We invite the scientific community to engage with this repository by submitting pull requests and opening issues for corrections, updating of values, or suggestions of new data that are relevant to the scope of this work.

## Acknowledgments

We are incredibly grateful for the generosity of a wide array of experts for their advice, suggestions, and criticism of this work. Specifically, we thank Suzy Beeler, Lars Bildsten, Justin Bois, Chris Bowler, Matthew Burgess, Ken Caldeira, Jörn Callies, Sean B. Carroll, Ibrahim Cissé, Joel Cohen, Michelle Dan, Bethany Ehlmann, Gidon Eshel, Paul Falkowski, Daniel Fisher, Thomas Frederikse, Neil Fromer, Eric Galbraith, Lea Goentoro, Evan Groover, John Grotzinger, Soichi Hirokawa, Greg Huber, Christina Hueschen, Bob Jaffe, Elizabeth Kolbert, Thomas Lecuit, Raphael Magarik, Jeff Marlow, Brad Marston, Jitu Mayor, Elliot Meyerowitz, Lisa Miller, Dianne Newman, Luke Oltrogge, Nigel Orme, Victoria Orphan, Marco Pasti, Pietro Perona, Noam Prywes, Stephen Quake, Hamza Raniwala, Manuel Razo-Mejia, Thomas Rosenbaum, Benjamin Rubin, Alex Rubinsteyn, Shyam Saladi, Tapio Schneider, Murali Sharma, Alon Shepon, Arthur Smith, Matthieu Talpe, Wati Taylor, Julie Theriot, Tadashi Tokieda, Cat Triandifillou, Sabah Ul-Hasan, Tine Valencic, and Ned Wingreen. We also thank Yue Qin for sharing data related to global water consumption. Many of the topics in this work began during the Applied Physics 150C course taught at Caltech during the early days of the COVID-19 pandemic. This work was supported by the Resnick Sustainability Institute at Caltech and the Schwartz-Reisman Collaborative Science Program at the Weizmann Institute of Science.

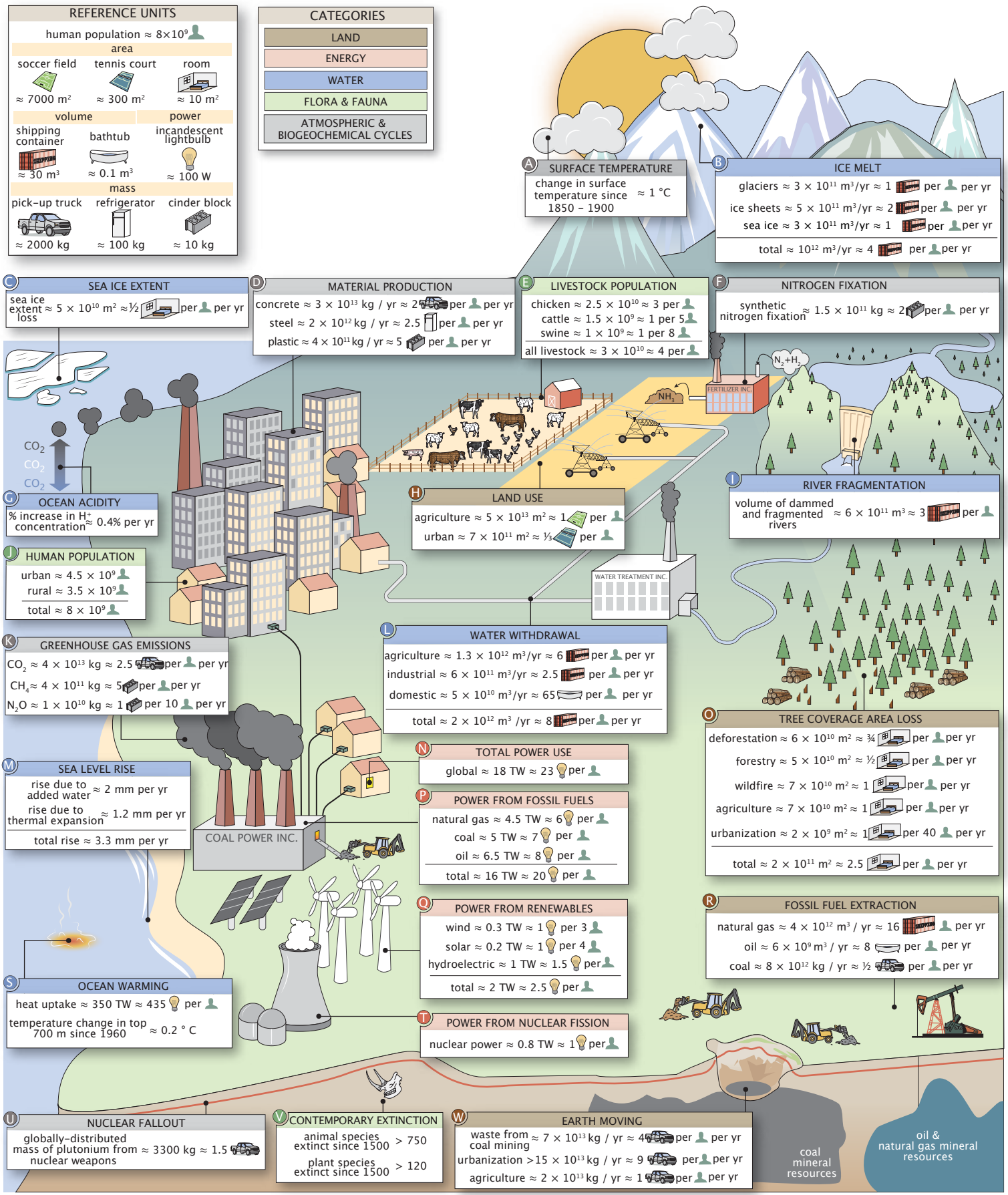

**Figure 1:** Human impacts on the planet and their relevant magnitudes. Relative units and the broad organizational categories are shown in the top-left panels. Source information and contextual comments for each subpanel are presented in the Supplemental Information.

(A)
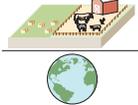
THE TERRA NUMBER

$$Te = \frac{\text{land area used by humans}}{\text{total land area of Earth}} \approx \frac{5 \times 10^7 \text{ km}^2}{15 \times 10^7 \text{ km}^2} \approx 0.3$$

(B)

THE DEFORESTATION NUMBER

$$Df = \frac{\text{annual forest loss from human action}}{\text{annual forest loss from wildfire}} \approx \frac{18 \times 10^4 \text{ km}^2/\text{yr}}{7 \times 10^4 \text{ km}^2/\text{yr}} \approx 2$$

(C)

THE BARNYARD NUMBER

$$By = \frac{\text{mass of terrestrial livestock}}{\text{mass of terrestrial wild animals}} \approx \frac{200 \times 10^{10} \text{ kg}}{7 \times 10^{10} \text{ kg}} \approx 30$$

(D)

THE NITROGEN NUMBER

$$N_2 = \frac{\text{mass of } N_2 \text{ fixed via the Haber-Bosch Process}}{\text{mass of } N_2 \text{ fixed biologically}} \approx \frac{1.5 \times 10^{11} \text{ kg/yr}}{1 \times 10^{11} \text{ kg/yr}} \approx 1$$

(E)

THE WATER NUMBER

$$Wa = \frac{\text{annual water volume used by humans}}{\text{global annual river discharge volume}} \approx \frac{2.2 \times 10^{12} \text{ m}^3/\text{yr}}{45 \times 10^{12} \text{ m}^3/\text{yr}} \approx 0.05$$

(F)
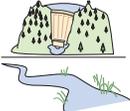
THE RIVER NUMBER

$$Rv = \frac{\text{river volume controlled by humans}}{\text{free-flowing river volume}} \approx \frac{6 \times 10^{11} \text{ m}^3}{6 \times 10^{11} \text{ m}^3} \approx 1$$

(G)
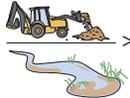
THE EARTH MOVER NUMBER

$$Em = \frac{\text{annual mass of earth moved by humans}}{\text{annual mass of earth moved by rivers}} \approx \frac{25 \times 10^{13} \text{ kg/yr}}{1.3 \times 10^{13} \text{ kg/yr}} > 15$$

(H)
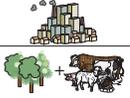
THE ANTHROPOMASS NUMBER

$$An = \frac{\text{total anthropomass}}{\text{total biomass}} \approx \frac{1.2 \times 10^{15} \text{ kg}}{1.2 \times 10^{15} \text{ kg}} \approx 1$$

(I)
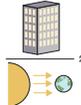
THE SOLAR NUMBER

$$Su = \frac{\text{annual human power usage}}{\text{annual incident solar power}} \approx \frac{20 \text{ TW}}{200{,}000 \text{ TW}} \approx 0.0001$$

(J)
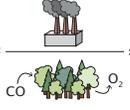
THE CO$_2$ NUMBER

$$CO_2 = \frac{\text{annual mass of anthropogenic CO}_2}{\text{annual mass of naturally removed CO}_2} \approx \frac{4 \times 10^{13} \text{ kg/yr}}{2 \times 10^{13} \text{ kg/yr}} \approx 2$$

(K)
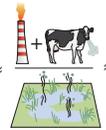
THE CH$_4$ NUMBER

$$Me = \frac{\text{annual mass of anthropogenic CH}_4}{\text{annual mass of natural CH}_4} \approx \frac{4 \times 10^{11} \text{ kg/yr}}{3 \times 10^{11} \text{ kg/yr}} \approx 1$$

(L)
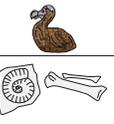
THE EXTINCTION NUMBER

$$Ex = \frac{\text{number of known animal extinctions}}{\text{number of expected natural animal extinctions}} = \frac{760 \text{ species}}{50 \text{ species}} > 10$$

**Figure 2:** Understanding human impacts through dimensionless numbers. The numerator for each number corresponds to the anthropogenic influence and the denominator corresponds to the value of the natural analogue.

# Supplemental Information for "The Anthropocene by the Numbers: A Quantitative Snapshot of Humanity's Influence on the Planet"

Griffin Chure[1,a], Rachel A. Banks[2], Avi I. Flamholz[2], Nicholas S. Sarai[3], Mason Kamb[4], Ignacio Lopez-Gomez[5], Yinon M. Bar-On[6], Ron Milo[6], Rob Phillips[2,4,7,*]

[1]Department of Applied Physics and Materials Science, California Institute of Technology, Pasadena, CA, USA
[2]Division of Biology and Biological Engineering, California Institute of Technology, Pasadena, CA, USA
[3]Division of Chemistry and Chemical Engineering, California Institute of Technology, Pasadena, CA, USA
[4]Chan-Zuckerberg BioHub, San Francisco, CA, USA
[5]Department of Environmental Science and Engineering, California Institute of Technology, Pasadena, CA, USA
[6]Department of Plant and Environmental Sciences, Weizmann Institute of Science, Rehovot, Israel
[7]Department of Physics, California Institute of Technology, Pasadena, CA, USA

[a]Current Address: Department of Biology, Stanford University, Stanford, CA, USA

# SI Section 1: References and Explanations For Values Reported in Figure 1

In this section, we report our extensive and detailed referencing for each and every quantity reported in the subpanels of Figure 1 of the main text. As described in the Materials & Methods, each value comes from the manual curation of a piece of scientific, industrial, governmental, or non-governmental organization reports, articles, or databases. Each value listed here contains information about the original source, the method used to obtain the value, as well as accession identification numbers for the Human Impacts Database (https://anthroponumbers.org), listed as HuIDs.

For each value, we attempt to provide an assessment of the uncertainty. For some values, this corresponds to the uncertainty in the measurement or inference as stated in the source material. In cases where a direct assessment of the uncertainty was not clearly presented, we sought other reported values for the same quantity from different data sources to present a range of the values. For others, this uncertainty represents the upper- and lower-bounds of the measurement or estimation.

Each value reported here is prefixed with a symbol representing our confidence in the value. Cases in which an equality (=) symbol is used represents that a measure of the uncertainty is reported in the original data source or represents a range of values from different sources that are tightly constrained (with 2 significant digits). An approximation symbol (≈) indicates values that we are confident in to within a factor of a few. In some cases, an approximation symbol (≈) represents a range where the values from different sources differ within three significant digits. In these cases, the ranges are presented as well. Finally, in some cases only a lower-bound for the quantity was able to be determined. These values are indicated by the use of an inequality symbol (>),.

## A. SURFACE WARMING

*Surface temperature change from the 1850-1900 average ≈ 1.0 - 1.4 °C  (HuID: 79598, 76539, 12147)*
**Data Source(s):** HadCRUT.4.6 (Morice et al., 2012, DOI: 10.1029/2011JD017187), GISTEMP v4 (GISTEMP Team, 2020: GISS Surface Temperature Analysis (GISTEMP), version 4. NASA Goddard Institute for Space Studies. Dataset accessed 2020-12-17 at https://data.giss.nasa.gov/gistemp/ & Lenssen et al., 2019, DOI: 10.1029/2018JD029522) and NOAAGlobalTemp v5 (Zhang et al, 2019, DOI: 10.1029/2019EO128229) datasets.
**Notes:** The global mean surface temperature captures near-surface air temperature over the planet's land and ocean surface. The value reported represents the spread of three estimates and their 95% confidence

intervals. Since data for the period 1850-1880 are missing in GISTEMP v4 and NOAAGlobalTemp v5, data are centered by setting the 1880-1900 mean of all datasets to the HadCRUT.4.6 mean over the same period.

## B. Annual Ice Melt

*Glaciers = (3.0 ± 1.2) × $10^{11}$ $m^3$ / yr (HuID: 32459)*
**Data Sources:** Intergovernmental Panel on Climate Change (IPCC) 2019 Special Report on the Ocean and Cryosphere in a Changing Climate. Table 2.A.1 on pp. 199-202.
**Notes:** Value corresponds to the trend of annual glacial ice volume loss (reported as ice mass loss) from major glacierized regions (2006-2015) based on aggregation of observation methods (original data source: Zemp et al. 2019, DOI:10.1038/s41586-019-1071-0) with satellite gravimetric observations (original data source: Wouters et al. 2019, DOI:10.3389/feart.2019.00096). Ice volume loss was calculated from ice mass loss assuming a standard pure ice density of 920 kg / $m^3$. Uncertainty represents a 95% confidence interval calculated from standard error propagation of the 95% confidence intervals reported in the original sources assuming them to be independent.

*Ice sheets = (4.7 ± 0.4) × $10^{11}$ $m^3$ / yr (HuIDs: 95798; 93137)*
**Data Source(s):** D. N. Wiese et al. 2019 JPL GRACE and GRACE-FO Mascon Ocean, Ice, and Hydrology Equivalent HDR Water Height RL06M CRI Filtered Version 2.0, Ver. 2.0, PO.DAAC, CA, USA. Dataset accessed [2020-Aug-10]. DOI: 10.5067/TEM- SC-3MJ62
**Notes:** Value corresponds to the trends of combined annual ice volume loss (reported as ice mass loss) from the Greenland and Antarctic Ice Sheets (2002-2020) measured by satellite gravimetry. Ice volume loss was calculated from ice mass loss assuming a standard pure ice density of 920 kg / $m^3$. Uncertainty represents one standard deviation and considers only propagation of monthly uncertainties in measurement.

*Arctic sea ice = (3.0 ± 1.0) × $10^{11}$ $m^3$ / yr (HuID: 89520)*
**Data Source(s):** PIOMAS Arctic Sea Ice Volume Reanalysis, Figure 1 of webpage as of October 31, 2020. Original method source: Schweiger et al. 2011, DOI:10.1029/2011JC007084
**Notes:** Value reported corresponds to the trend of annual volume loss from Arctic sea ice (1979-2020). The uncertainty in the trend represents the range in trends calculated from three ice volume determination methods.

## C. Sea Ice Extent

*Extent of loss at yearly maximum cover (September) ≈ 8.4 × $10^{10}$ $m^2$ / yr (HuID: 33993)*
*Extent loss at yearly minimum cover (March) ≈ 4.0 × $10^{10}$ $m^2$ / yr (HuID: 87741)*
*Average annual extent loss = 5.5 ± 0.2 × $10^{10}$ $m^2$ / yr (HuID: 70818)*
**Data Source(s):** Comiso et al. 2017, DOI:10.1002/2017JC012768. Fetterer et al. 2017, updated daily. Sea Ice Index, Version 3, Boulder, Colorado USA. NSIDC: National Snow and Ice Data Center, DOI:10.7265/N5K072F8, [Accessed 2020-Oct-19].
**Notes:** Sea ice extent refers to the area of the sea with > 15% ice coverage. Annual value corresponds to the linear trend of annually averaged Arctic sea ice extent from 1979-2015 (Comiso et al. 2017) calculated from four different methods. This is in good agreement with the linear trend of annual extent loss calculated by averaging over every month in a given year (5.5 × $10^{10}$ $m^2$ / yr HuID: 66277). The minimum cover extent loss corresponds to the linear trend of Arctic sea ice extent in September from 1979-2020 and the maximum cover extent loss corresponds to the linear trend of sea ice extent in March from 1979-2020. The Antarctic sea ice extent trend is not shown because a significant long-term trend over the satellite observation period is not observed and short-term trends are not yet identifiable.

## D. Annual Material Production

*Concrete production ≈ (2 - 3) × $10^{13}$ kg / yr (HuID: 25488; 81346; 16995)*

**Data Source(s):** United States Geological Survey (USGS), Mineral Commodity Summaries 2020, pp. 42-43, DOI:10.3133/mcs2020. Miller et al. 2016, Table 1, DOI:10.1088/1748-9326/11/7/074029. Monteiro et al. 2017, DOI:10.1038/nmat4930. Krausmann et al. 2017, DOI:10.1073/pnas.1613773114

**Notes:** Concrete is formed when aggregate material is bonded together by hydrated cement. The USGS reports the mass of cement produced in 2019 as $4.1 \times 10^{12}$ kg. As most cement is used to form concrete, cement production can be used to estimate concrete mass using a multiplicative conversion factor of 7 (Monteiro et al.). Miller et al. report that the cement, aggregate and water used in concrete in 2012 sum to $2.3 \times 10^{13}$ kg. Krausmann et al. report an estimated value from 2010 based on a material input, stocks, and outputs model. The value is net annual addition to concrete stocks plus annual waste and recycling to estimate gross production of concrete.

*Steel production = $(1.4 - 1.9) \times 10^{12}$ kg / yr (HuID: 51453; 44894; 85981)*
**Data Source(s):** United States Geological Survey (USGS), Mineral Commodity Summaries 2020, pp. 82-83, DOI:10.3133/mcs2020. World Steel Association, World Steel in Figures 2020, p. 6. Krausmann et al. 2017, DOI:10.1073/pnas.1613773114

**Notes:** Crude steel includes stainless steels, carbon steels, and other alloys. The USGS reports the mass of crude steel produced in 2019 as 1900 megatonnes (Mt). The World Steel Association reports a production value of 1869 Mt in 2019. Krausmann et al. report an estimated value from 2010 based on a material input, stocks, and outputs model. The value is net annual addition to steel stocks plus annual waste and recycling to estimate gross production of steel.

*Plastic production ≈ $4 \times 10^{11}$ kg / yr (HuID: 97241; 25437)*
**Data Source(s):** Geyer et al. 2017, Table S1, DOI:10.1126/sciadv.1700782. Krausmann et al. 2017, DOI:10.1073/pnas.1613773114

**Notes:** Value represents the approximate sum total global production of plastic fibers and plastic resin during the calendar year of 2015. Comprehensive data about global plastic production is sorely lacking. Geyer et al. draw data from various industry groups to estimate total production of different polymers and additives. Some of the underlying data is not publicly available, and data from financially-interested parties is inherently suspect. Krausmann et al. report an estimated value from 2010 based on a material input, stocks, and outputs model. The value is net annual addition to stocks plus annual waste and end-of-life recycling to estimate gross production of plastics.

## E. Livestock Population

*Chicken standing population ≈ $2.5 \times 10^{10}$ (HuID: 94934)*
*Cattle standing population ≈ $1.5 \times 10^{9}$ (HuID: 92006)*
*Swine standing population ≈ $1 \times 10^{9}$ (HuID: 21368)*
*All livestock standing population ≈ $3 \times 10^{10}$ (HuID: 43599)*
**Data Source(s):** Food and Agriculture Organization (FAO) of the United Nations Statistical Database (2020) — Live Animals.
**Notes:** Counts correspond to the estimated standing populations in 2018. Values are reported directly by countries. The FAO uses non-governmental statistical sources to address uncertainty and missing (non-reported) data. Reported values are therefore approximations.

## F. Annual Synthetic Nitrogen Fixation

*Annual mass of synthetically fixed nitrogen ≈ $1.5 \times 10^{11}$ kg N / yr (HuID: 60580; 61614)*
**Data Source(s):** United States Geological Survey (USGS), Mineral Commodity Summaries 2020, pp. 116-117, DOI:10.3133/mcs2020. International Fertilizer Association (IFA) Statistical Database (2020) — Ammonia Production & Trade Tables by Region. Smith et al. 2020, DOI: 10.1039/c9ee02873k.
**Notes:** Ammonia ($NH_3$) produced globally is compiled by the USGS and IFA from major factories that report output. The USGS estimates the approximate mass of nitrogen in ammonia produced in 2018 as $1.50 \times 10^{11}$ kg N and the International Fertilizer Association reports a production value of $1.50 \times 10^{11}$ kg N in 2019. Nearly all of this mass is produced by the Haber-Bosch process (>96%, Smith et al. 2020). In the United States most of this mass is used for

fertilizer, with the remainder being used to synthesize nitrogen-containing chemicals including explosives, plastics, and pharmaceuticals (≈ 88%, USGS Mineral Commodity Summaries 2020).

## G. Ocean Acidity

*Surface ocean [H+] ≈ 0.2 parts per billion* (HuID: 90472)
*Annual change in [H+] = 0.36 ± 0.03%* (HuID: 19394)
**Data Source(s):** Figures 1-2 of European Environment Agency report CLIM 043 (2020). Original data source of the report is "Global Mean Sea Water pH" from Copernicus Marine Environment Monitoring Service.
**Notes:** Reported value is calculated from the global average annual change in pH over years 1985-2018. The average oceanic surface pH was ≈ 8.057 in 2018 and decreases annually by ≈ 0.002 units, giving a change in [H+] of roughly $10^{-8.055} - 10^{-8.057} ≈ 4 \times 10^{-11}$ mol/L or about 0.4% of the global average. [H+] is calculated as $10^{-pH} ≈ 10^{-8}$ mol/L or 0.2 parts per billion (ppb), noting that $[H_2O] ≈ 55$ mol/L. Uncertainty for annual change is the standard error of the mean.

## H. Land Use

*Agriculture ≈ $5 \times 10^{13}$ m$^2$* (HuID: 29582)
**Data Source(s):** Food and Agriculture Organization (FAO) of the United Nations Statistical Database (2020) — Land Use.
**Notes:** Agricultural land is defined as all land that is under agricultural management including pastures, meadows, permanent crops, temporary crops, land under fallow, and land under agricultural structures (such as barns). Reported value corresponds to 2017 estimates by the FAO.
*Urban ≈ (6 - 8) $\times 10^{11}$ m$^2$* (HuID: 41339; 39341)
**Data Source(s):** Florczyk et al. 2019 (https://tinyurl.com/yyxxgtll) and Table 3 of Liu et al. 2018 DOI: 10.1016/j.rse.2018.02.055
**Notes:** Urban land area is determined from satellite imagery. An area is determined to be "urban" if the total population is greater than 5,000 and has a minimum population density of 300 people per km$^2$. Reported value gives the range of recent measurements of ≈ $6.5 \times 10^{11}$ m$^2$ (2015) and ≈ $(7.5 ± 1.5) \times 10^{11}$ m$^2$ (2010) from Florczyk et al. 2019 and Liu et al. 2018, respectively.

## I. River Fragmentation

*Global fragmented river volume ≈ $6 \times 10^{11}$ m$^3$* (HuID: 61661)
**Data Source(s):** Grill et al. 2019 DOI: 10.1038/s41586-019-1111-9
**Notes:** Value corresponds to the water volume contained in rivers that fall below the connectivity threshold required to classify them as free-flowing. Value considers only rivers with upstream catchment areas greater than 10 km$^2$ or discharge volumes greater than 0.1 m$^3$ per second. The ratio of global river volume in disrupted rivers to free-flowing rivers is approximately 0.9. The exact value depends on the cutoff used to define a "free-flowing" river. We direct the reader to the source for thorough detail.

## J. Human Population

*Urban population ≈ 55%* (HuID: 93995)
*Global population ≈ $7.6 \times 10^9$ people* (HuID: 85255)
**Data Source(s):** Food and Agricultural Organization (FAO) of the United Nations Report on Annual Population, 2019.
**Notes:** Value for total population in 2018 comes from a combination of direct population reports from country governments as well as inferences of underreported or missing data. The definition of "urban" differs between countries and the data does not distinguish between urban and suburban populations despite substantive differences

between these land uses (Jones & Kammen 2013, DOI: 10.1021/es4034364). As explained by the United Nations population division, "When the definition used in the latest census was not the same as in previous censuses, the data were adjusted whenever possible so as to maintain consistency." Rural population is computed from this fraction along with the total human population, implying that the total population is composed only of "urban" and "rural" communities.

## K. Greenhouse Gas Emissions

*Anthropogenic $CO_2$ = (4.25 ± 0.33) × $10^{13}$ kg $CO_2$ / yr (HuID: 24789; 54608; 98043; 60670)*
**Data Source(s):** Table 6 of Friedlingstein et al. 2019, DOI: 10.5194/essd-11-1783-2019. Original data sources relevant to this study compiled in Friedlingstein et al.: 1) Gilfillan et al. https://energy.appstate.edu/CDIAC 2) Average of two bookkeeping models: Houghton and Nassikas 2017 DOI: 10.1002/2016GB005546; Hansis et al. 2015 DOI: 10.1002/2014GB004997. 3) Dlugokencky and Tans, National Oceanic & Atmospheric Administration, Earth System Research Laboratory (NOAA/ESRL), https://www.esrl.noaa.gov/gmd/ccgg/trends/global.html, [Accessed 3-Nov-2019].
**Notes:** Value corresponds to total $CO_2$ emissions from fossil fuel combustion, industry (predominantly cement production), and land-use change during calendar year 2018. Emissions from land-use change are due to the burning or degradation of plant biomass. In 2018, roughly 1.88 × $10^{13}$ kg $CO_2$ / yr accumulated in the atmosphere, reflecting the balance of emissions and $CO_2$ uptake by plants and oceans (Dlugokencky and Tans). Uncertainty corresponds to one standard deviation.

*Anthropogenic $CH_4$ = (3.4 - 3.9) × $10^{11}$ kg $CH_4$ / yr (HuID: 96837; 30725)*
**Data Source(s):** Table 3 of Saunois, et al. 2020. DOI: 10.5194/essd-12-1561-2020.
**Notes:** Value corresponds to 2008-2017 decadal average mass of $CH_4$ emissions from anthropogenic sources. Includes emissions from agriculture and landfill, fossil fuels, and burning of biomass and biofuels, but other inventories of anthropogenic methane emissions are also considered. Reported range represents the minimum and maximum estimated emissions from a combination of "bottom-up" and "top-down" models.

*Anthropogenic $N_2O$ = 1.1 (+0.6, -0.5) × $10^{10}$ kg $N_2O$ / yr (HuID: 44575)*
**Data Source(s):** Table 1 of Tian, H., et al. 2020. DOI: 10.1038/s41586-020-2780-0
**Notes:** Value corresponds to annualized $N_2O$ emissions from anthropogenic sources in the years 2007-2016. The value reported in the source is 7.3 [4.2, 11.4] Tg N / year. This is converted to a mass of $N_2O$ using the fact that N ≈ 14/22 of the mass of $N_2O$. Reported value is mean with the uncertainty bounds (+,-) representing the maximum and minimum values observed in the 2007-2016 time period.

## L. Water Withdrawal

*Agricultural = 1.3 × $10^{12}$ $m^3$ / year (HuID: 84545, 43593, 95345)*
*Industrial = 5.9 × $10^{11}$ $m^3$ / year (HuID: 27142)*
*Domestic = 5.4 × $10^{10}$ $m^3$ / year (HuID: 69424)*
*Total = (1.7 - 2.2) × $10^{12}$ $m^3$ / year (HuID: 27342, 68004)*
**Data Source(s):** Figure 1 of Qin et al. 2019. DOI: 10.1038/s41893-019-0294-2. AQUASTAT Main Database, Food and Agriculture Organization of the United Nations
**Notes:** "Agricultural" and "total" withdrawal include one value from Qin et al. (who reports "consumption") and one value from the AQUASTAT database. Industrial water withdrawal is from AQUASTAT and domestic withdrawal value is from Qin et al. Values in AQUASTAT are self-reported by countries and have missing values from some countries, probably accounting for a few percent underreporting. All values represent water withdrawals. For agricultural and domestic, water withdrawal is assumed to be the same as water consumption, which is reported in Qin et al.

## M. Sea Level Rise

*Added water = 1.97 (+0.36, -0.34) mm / yr (HuID: 97108)*

*Thermal expansion = 1.19 (+0.25, -0.24) mm / yr (HuID: 97688)*
*Total observed sea-level rise = 3.35 (+0.47, -0.44) mm / yr (HuID: 81373)*
**Data Source(s):** Table 1 of Frederikse et al. 2020. DOI:10.1038/s41586-020-2591-3.
**Notes:** Values correspond to the average global sea level rise for the years 1993 - 2018. "Added water" (barystatic) change includes effects from meltwater from glaciers and ice sheets, added mass from sea-ice discharge, and changes in the amount of terrestrial water storage. Thermal expansion accounts for the volume change of water with increasing temperature. Values for "thermal expansion" and "added water" come from direct observations of ocean temperature and gravimetry/altimetry, respectively. Total sea level rise is the observed value using a combination of measurement methods. "Other sources" reported in Figure 1 accounts for observed residual sea level rise not attributed to a source in the model. Values in brackets correspond to the upper and lower bounds of the 90% confidence interval.

## N. Total Power Use

*Global power use ≈ 19 - 20 TW (HuID: 31373; 85317)*
**Data Source(s):** bp Statistical Review of World Energy, 2020; U.S. Energy Information Administration, 2020.
**Notes:** Value represents the sum of total primary energy consumed from oil, natural gas, coal, and nuclear energy and electricity generated by hydroelectric and other renewables. Value is calculated using annual primary energy consumption as reported in data sources assuming uniform use throughout a year, yielding ≈ 19 - 20 TW.

## O. Tree Coverage Area Loss

*Commodity-driven deforestation = (5.7 ± 1.1) × $10^{10}$ $m^2$ / yr (HuID: 96098)*
*Forestry = (5.4 ± 0.8) × $10^{10}$ $m^2$ / yr (HuID: 38352)*
*Urbanization = (2 ± 1) × $10^9$ $m^2$ / yr (HuID: 19429)*
*Shifting agriculture = (7.5 ± 0.9) × $10^{10}$ $m^2$ / yr (HuID: 24388)*
*Wildfire = (7.2 ± 1.3) × $10^{10}$ $m^2$ / yr (HuID: 92221)*
*Total tree cover area loss ≈ 2 × $10^{11}$ $m^2$ / yr (HuID: 78576)*
**Data Source(s):** Table 1 of Curtis et al. 2018 DOI:10.1126/science.aau3445. Hansen et al. 2013 DOI:10.1126/science.1244693. Global Forest Watch, 2020. Reported values in source correspond to total loss from 2001 - 2015. Values given are averages over this 15 year window.
**Notes:** Commodity-driven deforestation is "long-term, permanent, conversion of forest and shrubland to a non-forest land use such as agriculture, mining, or energy infrastructure." Forestry is defined as large-scale operations occurring within managed forests and tree plantations with evidence of forest regrowth in subsequent years. Urbanization converts forest and shrubland for the expansion and intensification of existing urban centers. Disruption due to "shifting agriculture" is defined as "small- to medium-scale forest and shrubland conversion for agriculture that is later abandoned and followed by subsequent forest regrowth". Disruption due to wildfire is "large-scale forest loss resulting from the burning of forest vegetation with no visible human conversion or agricultural activity afterward." Uncertainty corresponds to the reported 95% confidence interval. Uncertainty is approximate for "urbanization" as the source reports an ambiguous error of "± <1%."

## P. Power From Fossil Fuels

*Natural gas = 4.5 - 4.8 TW (HuID: 49947; 86175)*
*Oil = 6.1 - 6.6 TW (HuID: 42121; 39756)*
*Coal = 5.0 - 5.5 TW (HuID: 10400; 60490)*
*Total = 16 - 17.0 TW (HuID: 29470; 29109 )*
**Data Source(s):** bp Statistical Review of World Energy, 2020. U.S. Energy Information Administration, 2020.
**Notes:** Values are self-reported by countries. All values from bp Statistical Review correspond to 2019 whereas values from the EIA correspond to 2018 estimates. Reported TW values are computed from primary energy units (e.g. kg coal) assuming uniform use throughout the year. Oil volume includes crude oil, shale oil, oil sands,

## Q. Power From Renewable Resources

*Wind = 0.36 - 0.39 TW (HuID: 30581, 85919)*
*Solar = 0.18 - 0.20TW (HuID: 99885, 58303)*
*Hydroelectric = 1.2 - 1.3 TW (HuID: 15765, 50558)*
*Total = 1.9 - 2.1 TW (HuID: 74571, 20246)*
**Data Source(s):** bp Statistical Review of World Energy, 2020. U.S. Energy Information Administration, 2020.
**Notes:** Reported values correspond to estimates for the 2019 calendar year for BP and 2018 for EIA data, except for total renewables, which is from 2017. Renewable resources are defined as wind, geothermal, solar, biomass and waste. Hydroelectric, while presented here, is not defined as a renewable in the BP dataset. All values are reported as input-equivalent energy, meaning the input energy that would have been required if the power was produced by fossil fuels. BP reports that fossil fuel efficiency used to make this conversion was about 40% in 2017.

## R. Fossil Fuel Extraction

*Natural gas volume = (3.9 - 4.0) × $10^{12}$ $m^3$ / yr (HuID: 11468; 20532)*
*Oil volume = (5.5 - 5.8) × $10^9$ $m^3$ / yr (HuID: 66789; 97719)*
*Coal mass = (7.8 - 8.1) × $10^{12}$ kg / yr (HuID: 78435; 48928)*
**Data Source(s):** bp Statistical Review of World Energy, 2020. U.S. Energy Information Administration (EIA), 2020.
**Notes:** Oil volume includes crude oil, shale oil, oil sands, condensates, and natural gas liquids separate from specific natural gas mining. Natural gas value excludes gas flared or recycled and includes natural gas produced for gas-to-liquids transformation. Coal value includes solid commercial fuels such as bituminous coal, anthracite, lignite, subbituminous coal, and other solid fuels. All values from bp Statistical Review correspond to 2019 whereas values from the EIA correspond to 2018 estimates.

## S. Ocean Warming

*Heat uptake = $346 \pm 51\ TW$ (HuID: 94108)*
*Upper ocean (0 - 700m) temperature increase since to 1960 = $0.18 - 0.20$ °C (HuID: 69674, 72086)*
**Data Source(s):** Table S1 of Cheng et al. 2017. DOI: 10.1126/sciadv.1601545. NOAA National Centers for Environmental Information, 2020. DOI: 10.1029/2012GL051106.
**Notes:** Heat uptake reported is the average over time period 1992-2015 with 95% confidence intervals. Range of temperatures reported captures the 95% confidence interval of temperature increase for the period 2015-2019 with respect to the 1958-1962 mean. Temperature change is considered in the upper 700 m because sea surface temperatures have high decadal variability and are a poor indicator of ocean warming; see Roemmich et al. 2015, DOI: 10.1038/NCLIMATE2513.

## T. Power From Nuclear Fission

*Nuclear power ≈ 0.79 - 0.89 TW (HuID: 48387; 71725)*
**Data Source(s):** bp Statistical Review of World Energy, 2020. U.S. Energy Information Administration (EIA), 2020
**Notes:** Values are self-reported by countries and correspond to estimates for the 2019 calendar year from BP and 2018 from EIA. Values are reported as 'input-equivalent' energy, meaning the energy that would have been needed to produce a given amount of power if the input were a fossil fuel, which is converted to TW here. This is calculated by multiplying the given power by a conversion factor representing the efficiency of power production by fossil fuels. In 2017, this factor was about 40%.

## U. Nuclear Fallout

*Anthropogenic $^{239}$Pu and $^{240}$Pu from nuclear weapons ≈ 1.4 $\times$ 10$^{11}$ kg / yr  (HuID: 42526)*

**Data Source(s):** Table 1 in Hancock et al. 2014 doi: 10.1144/SP395.15. Fallout in activity from UNSCEAR 2000 Report on Sources and Effects of Ionizing Radiation Report to the UN General Assembly -- Volume 1.

**Notes:** The approximate mass of Plutonium isotopes $^{239}$Pu and $^{240}$Pu released into the atmosphere from the ≈ 500 above-ground nuclear weapons tests conducted between 1945 and 1980. Naturally occurring $^{239}$Pu and $^{240}$Pu are rare, meaning that nearly all contemporary labile plutonium comes from human production (Taylor 2001,doi: 10.1016/S1569-4860(01)80003-6). The total mass of radionuclides released is ≈ 3300 kg with a combined radioactive fallout of ≈ 11 PBq. These values do not represent the entire 239+240Pu globally distributed mass as it excludes non-weapons sources.

## V. Contemporary Extinction

*Animal species extinct since 1500  > 750 (HuID: 44641)*
*Plant species extinct since 1500  > 120 (HuID: 86866)*

**Data Source(s):** The IUCN Red List of Threatened Species. Version 2020-2

**Notes:** Values correspond to absolute lower-bound count of animal extinctions over the past ≈ 520 years. Of the predicted ≈ 8 million animal species, the IUCN databases catalogues only ≈ 900,000 with only ≈ 75,000 being assigned a conservation status. Representation of plants and fungi is even more sparse with only ≈ 40,000 and ≈ 285 being assigned a conservation status, respectively. The number of extinct animal species is undoubtedly higher than these reported values, as signified by an inequality symbol (>).

## W. Earth Moving

*Waste and overburden from coal mining ≈ 6.5 × 10$^{13}$ kg / yr (HuID: 72899)*
*Earth moved from urbanization  > 1.4  × 10$^{14}$ kg / yr (HuID: 59640)*

**Data Source(s):** Supplementary table 1 of Cooper et al. 2018. DOI: doi.org/gfwfhd.

**Notes:** Coal mining waste and overburden mass is calculated given commodity-level stripping ratios (mass of overburden/waste per mass of coal resource mined) and reported values of global coal production by type. Urbanization mass is presented as a lower bound estimate of the mass of earth moved from global construction projects. This comes from a conservative estimate that the ratio of the mass of earth moved per mass of cement/concrete used in construction globally is 2:1. This value is highly context dependent and we encourage the reader to read the source material for a more thorough description of this estimation.

*Erosion rate from agriculture > (1.2 - 2.4) × 10$^{13}$ kg / yr (HuID: 19415; 41496)*

**Data Source(s):** Pg. 377 of Wang and Van Oost 2019. DOI: 10.1177/0959683618816499. Pg.  21996 of Borrelli et al. 2020 DOI: 10.1073/pnas.2001403117.

**Notes:** Cumulative sediment mass loss over history of human agriculture due to accelerated erosion is estimated to be ≈ 30,000 Gt. Recent years have an estimated erosion rate ranging from  12 Pg / yr (Wang and Van Oost) to ≈ 24 Pg / yr (Borrelli et al.). Values come from computational models conditioned on time-resolved measurements of sediment deposition in catchment basins.

# SI Section 2: Connections Between Panels in Figure 1

In the main text, we presented a few examples of connections between the values displayed in Figure 1 could be drawn. Another way to approach these values is to begin with the question "how much water do we use?" A vast majority of human water use is for agriculture and industrial use (including cooling power plants, Figure 1 K; HuID: 84545, 43593, 95345; 27142). In order to harness this water, about half of the world's river volume is now under human control (Figure 1 H; HuID: 61661). Agricultural water use is mainly for irrigation of crops; approximately 10% of the total agricultural land area is irrigated (≈ 5 × 10$^{12}$ m$^2$). These crops also require nitrogenous fertilizer to grow, requiring synthetic fixation of nitrogen (Figure 1 E; HuID: 60580, 61614). Nitrogen fixation as well as rice paddies and livestock are major sources of N$_2$O and CH$_4$ emissions (Figure 1 J; HuID: 96837, 30725; 44575). Another major use of water by humans is for cooling power plants that generate power from fossil fuels (Figure 1 O; HuID: 29470, 29109). Generating this power requires extraction of massive amounts of coal, oil, and natural gas (Figure 1 Q; HuID: 11468, 20532; 66789, 97719; 78435, 48928) and the movement of large amounts of geological materials (Figure 1 V; HuID: 72899).

## SI Section 3: Region Definitions

See the supplemental file "region definitions" for a list of countries and their associated regions used in this study. For tree cover area loss, we did not have access to data at the country level and used slightly different region definitions: Central & South America; North America; Russia, China & South Asia; Southeast Asia; Europe (except Russia); Africa; and Oceania

## SI Section 4: Discussion of Regional Distributions

While Figure 1 presents the magnitude of human impacts at a global scale, it is important to recognize that these human impacts — both their origins and repercussions — are highly variable across the globe. The distribution of the global population and the societal and cultural differences which prescribe our interactions with the planet lead to unequal contributions to these impacts featured in Figure 1. The outcomes of these impacts are also unequally distributed, leading to some regions being disproportionately affected by the consequences of human activities. Figure S1 displays a coarse regional breakdown of the numbers from Figure 1 for which regional distributions could be determined. The region definitions used in Figure S1 are similar to the definitions set forth by the Food and Agricultural Organization (FAO) of the United Nations, assigning the semi-continental regions of North America, South America, Africa, Europe, Asia, and Oceania. Supplementary Information Section 3 provides a list of the defined regions along with the countries and localities which form them. Here, we specify both the total contribution of each region and the per capita value given the population of that region at the year(s) in which the quantity was measured. While we have chosen here to present a somewhat simplistic regional breakdown, there are many possible ways to look at the data, such as by country or regional economic activity. We hope this resource inspires others to examine these data with different regional definitions.

Much as in the case of Figure 1, interesting details emerge naturally from the display of the data shown in Figure S1. For example, Asia dominates global agricultural water withdrawal, using about 62%, while Northern America takes the lead for industrial water withdrawal, much of which is used for the production of electricity. From considering the volume of water withdrawn per capita, however, we find that Northern America withdraws more water per person than any other region for agricultural, industrial, and domestic water use.

Northern America also emits far more CO$_2$ per person than any other region, with Oceania and Europe coming second and third, respectively. This disparity can be partially understood by considering how each region uses nuclear fission, fossil fuel combustion, and renewable resources as sources of energy. While Asia consumes half of total power, per capita consumption is markedly lower than North America, Europe, and Oceania due to Asia having more than fivefold greater population than those other regions. Interestingly, renewables and nuclear power tell a different story. Southern America, while consuming merely 4% of total power, generates about 14% of renewable energy. Nuclear power generation, on the other hand, is dominated by Northern America and Europe, while Oceania, which has only a single research-grade nuclear reactor, comes dead last.

Investigating forest loss by region and cause provides a clearer picture of the contemporary trends. At a global level, all drivers of forest loss are comparable in magnitude except for urbanization, which is responsible for ≈ 1% of total tree cover area loss. However, when each driver is broken down by region, it becomes apparent that not all regions are comparable. Central and South America account for 64% of commodity-driven deforestation (meaning, clearcutting with no substantial regrowth of tree cover), whereas a majority of forest loss due to shifting agriculture occurs in Africa (where regrowth does occur). Together, wildfires in North America, Russia, China, and South Asia make up nearly 90% of the fire-based loss in tree cover. North America alone, which has had periodic and enormous wildfires over the past two decades, accounts for around 42% of fire-based loss. Finally, urbanization is dominated by development in South Asia. It is important to realize that while urbanization at a global level is the smallest driver of tree cover loss, it can still have strong impacts at the regional level, greatly perturbing local ecosystems and biodiversity.

## THE GEOGRAPHY OF HUMAN IMPACTS

Figure 1 of the main text represents the impact humans have on the Earth at a global scale. While these numbers are handy, it is important to acknowledge that they vary from country-to-country and continent-to-continent. Furthermore, the consequences of these anthropogenic impacts are also unequally distributed, meaning some regions experience effects disproportionate to their contribution. Here, we give a sense of the geographic distribution of several values presented in Figure 1, broken down by continental region as shown below.

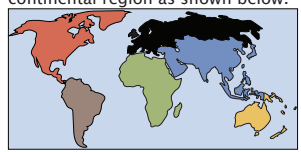

- Asia — (As)
- North America — (NA)
- South America — (SA)
- Europe — (Eu)
- Oceania — (Oc)
- Africa — (Af)

## THE LIVESTOCK POPULATION

The global population of terrestrial livestock is around 30 billion individuals, most of which are chickens. Asia houses most of the global livestock population, though South America and Europe harbor more animals on a per–capita basis.

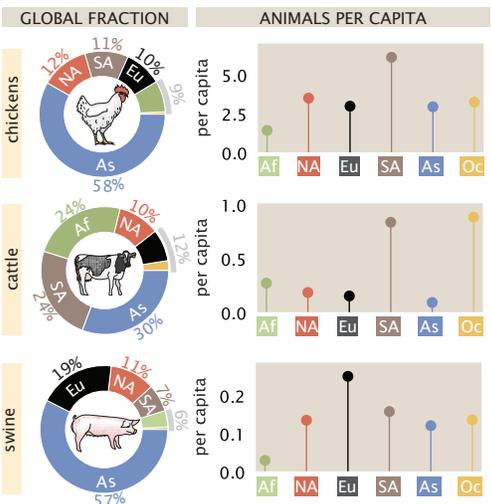

**Sources:** Food and Agricultural Organization of the United Nations

## NITROGENOUS FERTILIZER USE & PRODUCTION

Modern agriculture requires nitrogen in amounts beyond what is produced naturally. Asia synthesizes and consumes a large majority of fixed nitrogen. However, Europe and North America dominate per capita synthesis whereas Oceania consumes more fertilizer per capita than any other region.

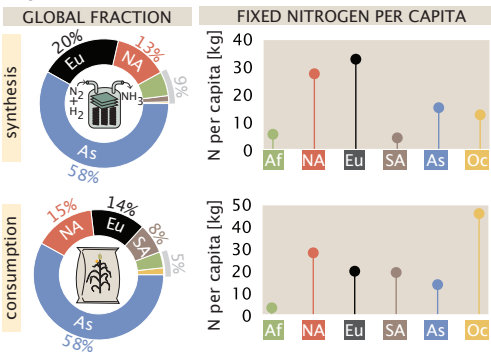

**Source:** Food and Agricultural Organization (FAO) of the United Nations.
**Notes:** Values account for reactive nitrogen production/consumption in context of fertilizer only and does not account for plastics, explosives, or other uses.

## THE HUMAN POPULATION

There are ≈ 8 billion humans on the planet, with approximately 50% living in 'urban' environments. The majority of the worlds population (as well as the majority of both urban and rural dwellers) live in Asia.

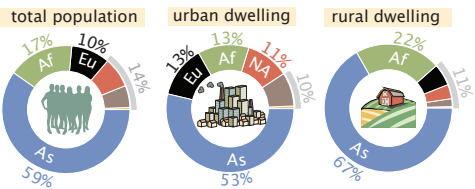

**Sources:** Food and Agricultural Organization of the United Nations – World Population
**Notes:** Urban/rural designation has no set definition and follows the conventions set by each reporting country.

## WATER WITHDRAWAL

While Asia withdraws the most water for agricultural and municipal needs, North America withdraws the plurality of water for industrial purposes. North America also withdraws more water per capita than any other region.

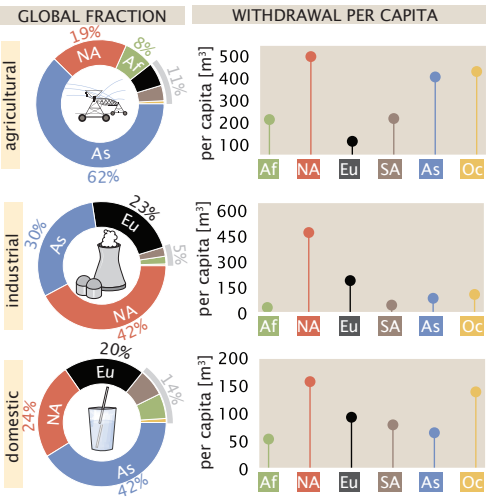

**Source:** AQUASTAT Main Database, Food and Agriculture Organization of the United Nations.
**Notes:** Values are reported directly from member countries and represent average of 2013–2017 period. Per capita values are computed using population of reporting countries.

## GREENHOUSE GAS EMISSIONS

$CO_2$ and $CH_4$ are two potent greenhouse gases which are routinely emitted by anthropogenic processes such as burning fuel and rearing livestock. While Asia emits roughly half of all $CO_2$ and $CH_4$, North America and Oceania produce the most on a per capita basis, respectively.

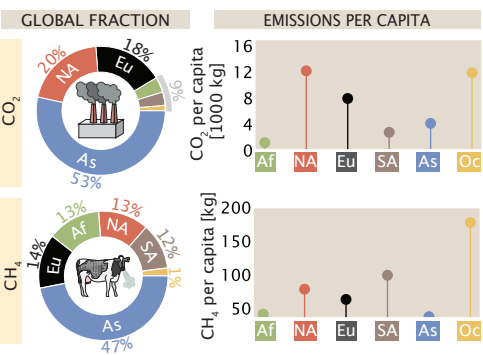

**Sources:** $CO_2$ data collated by: Friedlingstein, P. et al. (2019). doi: 10.5194/essd-11-1783-2019. See Panel J on Pg. 4 for complete list of sources. $CH_4$ data from Saunois et al, 2020 doi: 10.5194/essd-12-1561-2020 **Notes:** Values report decadal averages in kg $CO_2$ or $CH_4$ per year over time period 2008–2017.

## LAND USE

Though humans are nearly evenly split between urban and rural environments, agricultural land is the far more common use of land area. Together, Asia and Africa contain more than half of global agricultural land. Asia alone accomodates more than half of the global urban land area.

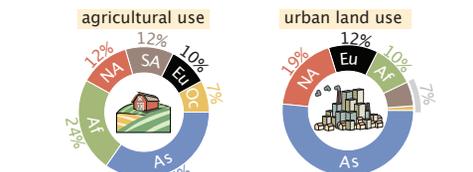

**Sources:** Food and Agricultural Organization (FAO) of the United Nations (2015) — Land Use [agricultural area]. Florczyk et al. 2019 — GHS Urban Centre Database 2015 [urban land area] **Notes:** Urban is defined as any inhabited area with ≥ 2500 residents, as defined by the USDA.

## TREE COVERAGE AREA LOSS

Most drivers of tree coverage area loss are comparable in their effect at a global scale. However, there are drastic regional differences in the relative magnitudes.

### REGION DEFINITION

- Central & South America
- North America
- Africa
- Russia, China, & South Asia
- Southeast Asia
- Europe (– Russia)
- Oceania

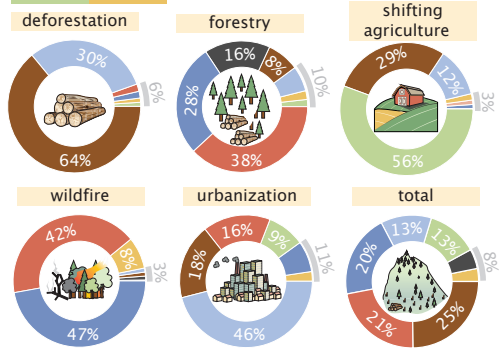

**Source:** Curtis et al. 2018 doi: 10.1126/science.aau3445.
**Notes:** Regions are as reported in Curtis et al. 2018. "Deforestation" here denotes permanent removal of tree cover for commodity production. "Shifting agriculture" here denotes forest/shrub land converted to agriculture and later abandoned. All values correspond to breakdown of cumulative tree cover area loss from 2001 – 2015.

## MATERIAL PRODUCTION

Humans excavate an enormous amount of material from the Earth's crust and transform it to build our structures. Two of these materials, concrete and steel, are produced primarily in Asia on both a global and per capita basis. Asia's per capita production of steel is only outpaced by Europe.

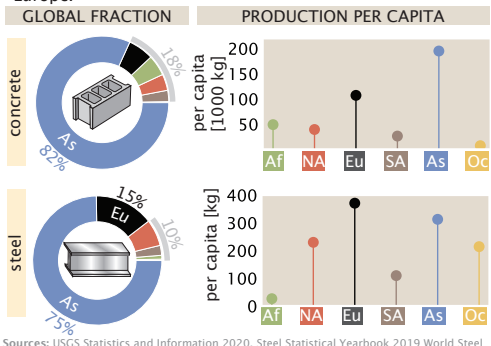

**Sources:** USGS Statistics and Information 2020, Steel Statistical Yearbook 2019 World Steel Association. Food and Agricultural Organization (FAO) of the United Nations — Annual Population. **Notes:** Reported values for cement and steel production corresponds to 2017 and 2018 values, respectively. Mass of concrete was calculated using a rule-of-thumb that 1 kg of cement yields 7 kg of concrete (Monteiro et al. 2017. doi: 0.138/nmat4930).

## POWER GENERATION AND CONSUMPTION

From heating water, to powering lights, to moving our vehicles, nearly every facet of modern human life requires the consumption of power, culminating in nearly 20 TW of power use in recent years. Asia consumes over half of the power derived from combustion of fossil fuels, with Europe and North America each consuming around 20% of the global total. Asia also produces the plurality of power from renewable technologies, such as hydroelectric, wind, and solar, however, North America, South America, and Europe each produce more on a per capita basis. Nuclear energy, however, is primarily produced in Europe, with North America and Asia coming in second and third place, respectively. On a per-capita basis, North America consumes or produces more energy than all other regions considered here, yielding a total power consumption of nearly 10,000 W per person.

**Source:** Energy Information Administration of the United States (2017)
**Notes:** "Renewables" includes hydroelectric, biofuels, biomass (wood), geothermal, wind, and solar. "Fossil fuels" includes coal, oil, and natural gas.

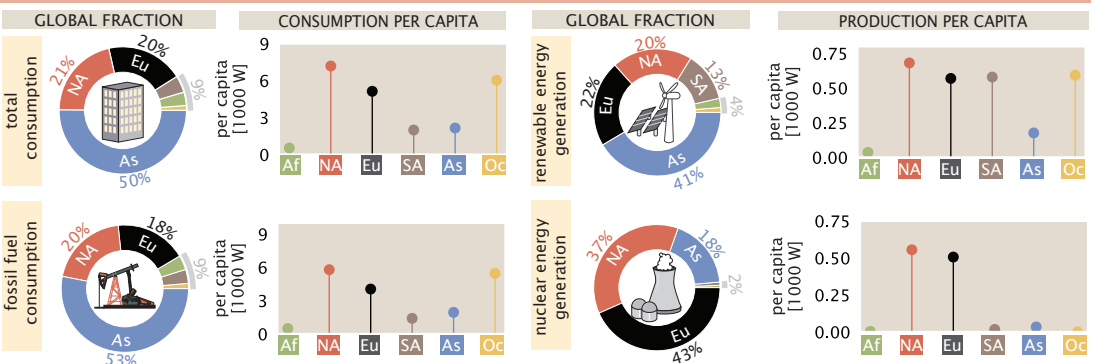

**Figure S1:** The regional distribution of human impacts on the planet. A subset of the quantities presented in Figure 1 are shown here broken down by geographic region. Unless otherwise noted, regions follow the Food and Agricultural Organization delineations of world regions, with the Caribbean subsumed into output for Northern America.

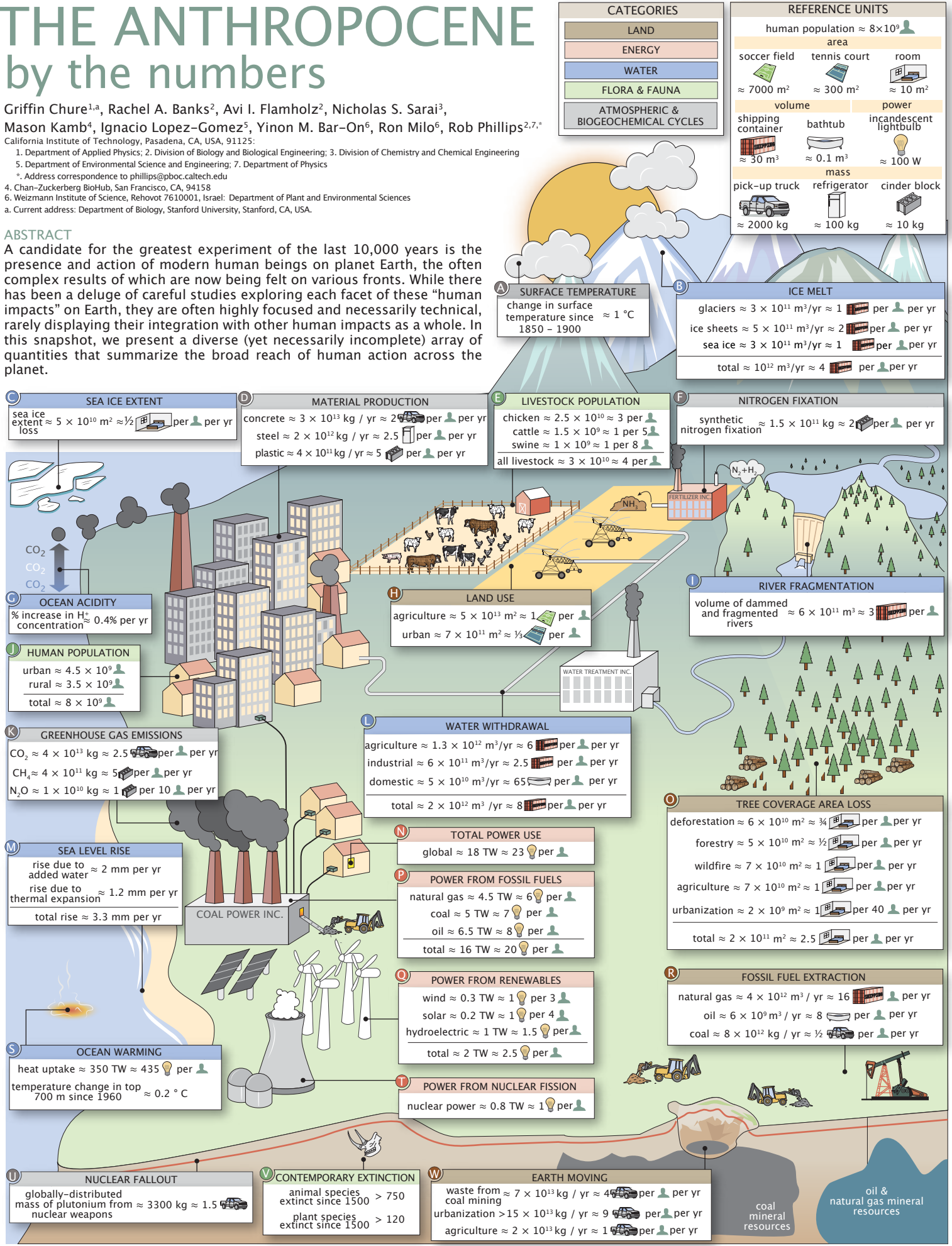

# The Anthropocene by the Numbers — Impacts By Region

## THE GEOGRAPHY OF HUMAN IMPACTS

Page 1 represents the impact humans have on the Earth at a global scale. While these numbers are handy, it is important to acknowledge that they vary from country-to-country and continent-to-continent. Furthermore, the consequences of these anthropogenic impacts are also unequally distributed, meaning some regions experience effects disproportionate to their contribution. Here, we give a sense of the geographic distribution of several values presented on page 1, broken down by continental region as shown below.

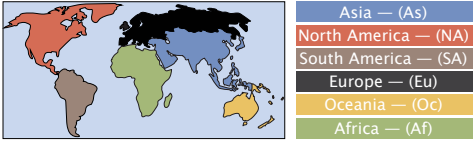

- Asia — (As)
- North America — (NA)
- South America — (SA)
- Europe — (Eu)
- Oceania — (Oc)
- Africa — (Af)

## THE LIVESTOCK POPULATION

The global population of terrestrial livestock is around 30 billion individuals, most of which are chickens. Asia houses most of the global livestock population, though South America and Europe harbor more animals on a per-capita basis.

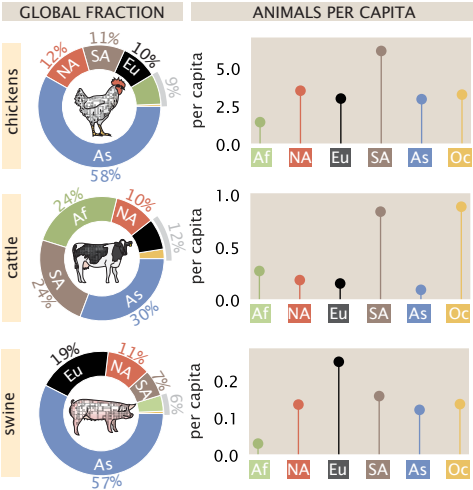

**Sources:** Food and Agricultural Organization of the United Nations

## NITROGENOUS FERTILIZER USE & PRODUCTION

Modern agriculture requires nitrogen in amounts beyond what is produced naturally. Asia synthesizes and consumes a large majority of fixed nitrogen. However, Europe and North America dominate per capita synthesis whereas Oceania consumes more fertilizer per capita than any other region.

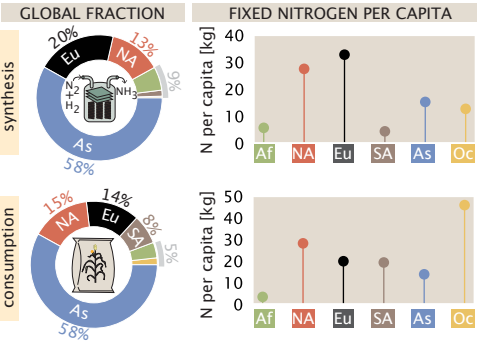

**Source:** Food and Agricultural Organization (FAO) of the United Nations.
**Notes:** Values account for reactive nitrogen production/consumption in context of fertilizer only and does not account for plastics, explosives, or other uses.

## THE HUMAN POPULATION

There are ≈ 8 billion humans on the planet, with approximately 50% living in 'urban' environments. The majority of the worlds population (as well as the majority of both urban and rural dwellers) live in Asia.

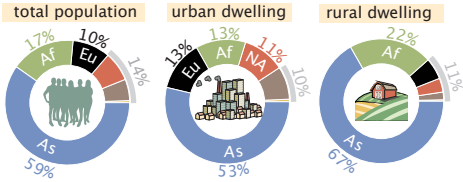

**Sources:** Food and Agricultural Organization of the United Nations – World Population
**Notes:** Urban/rural designation has no set definition and follows the conventions set by each reporting country.

## WATER WITHDRAWAL

While Asia withdraws the most water for agricultural and municipal needs, North America withdraws the plurality of water for industrial purposes. North America also withdraws more water per capita than any other region.

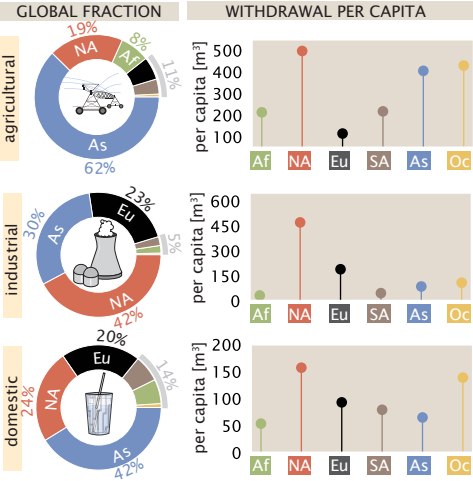

**Source:** AQUASTAT Main Database, Food and Agriculture Organization of the United Nations.
**Notes:** Values are reported directly from member countries and represent average of 2013–2017 period. Per capita values are computed given population of reporting countries.

## GREENHOUSE GAS EMISSIONS

$CO_2$ and $CH_4$ are two potent greenhouse gases which are routinely emitted by anthropogenic processes such as burning fuel and rearing livestock. While Asia emits roughly half of all $CO_2$ and $CH_4$, North America and Oceania produce the most on a per capita basis, respectively.

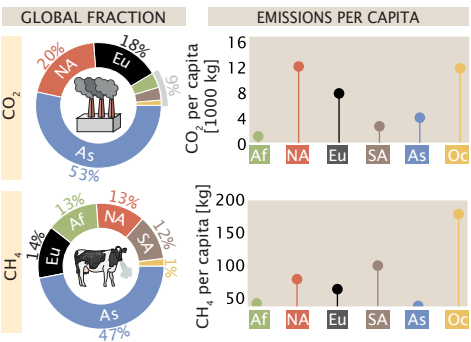

**Sources:** $CO_2$ data collated by: Friedlingstein, P. et al. (2019). doi: 10.5194/essd-11-1783-2019. See Panel J on Pg. 4 for complete list of sources. $CH_4$ data from Saunois et al, 2020 doi: 10.5194/essd-12-1561-2020 **Notes:** Values report decadal averages in kg $CO_2$ or $CH_4$ per year over time period 2008–2017.

## LAND USE

Though humans are nearly evenly split between urban and rural environments, agricultural land is the far more common use of land area. Together, Asia and Africa contain more than half of global agricultural land. Asia alone accomodates more than half of the global urban land area.

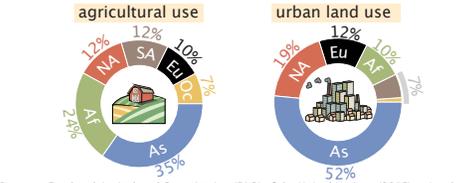

**Sources:** Food and Agricultural Organization (FAO) of the United Nations (2015) — Land Use [agricultural area]. Florczyk et al. 2019 — GHS Urban Centre Database 2015 [urban land area] **Notes:** Urban is defined as any inhabited area with ≥ 2500 residents, as defined by the USDA.

## TREE COVERAGE AREA LOSS

Most drivers of tree coverage area loss are comparable in their effect at a global scale. However, there are drastic regional differences in the relative magnitudes.

### REGION DEFINITION

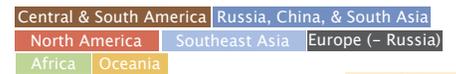

- Central & South America
- Russia, China, & South Asia
- North America
- Southeast Asia
- Europe (– Russia)
- Africa
- Oceania

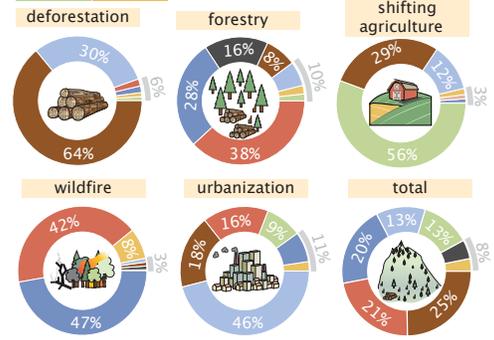

**Source:** Curtis et al. 2018 doi: 10.1126/science.aau3445.
**Notes:** Regions are as reported in Curtis et al. 2018. "Deforestation" here denotes permanent removal of tree cover for commodity production. "Shifting agriculture" here denotes forest/shrub land converted to agriculture and later abandoned. All values correspond to breakdown of cumulative tree cover area loss from 2001 – 2015.

## MATERIAL PRODUCTION

Humans excavate an enormous amount of material from the Earth's crust and transform it to build our structures. Two of these materials, concrete and steel, are produced primarily in Asia on both a global and per capita basis. Asia's per capita production of steel is only outpaced by Europe.

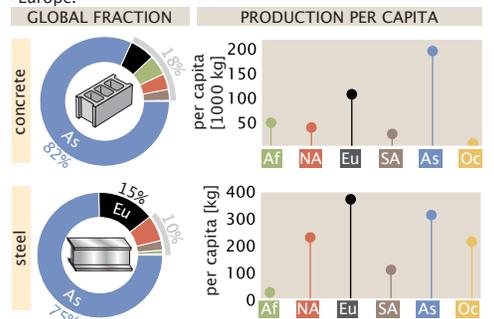

**Sources:** USGS Statistics and Information 2020, Steel Statistical Yearbook 2019 World Steel Association. Food and Agricultural Organization (FAO) of the United Nations — Annual Population. **Notes:** Reported values for cement and steel production corresponds to 2017 and 2018 values, respectively. Mass of concrete was calculated using a rule-of-thumb that 1 kg of cement yields 7 kg of concrete (Monteiro et al. 2017. doi: 0.138/nmat4930).

## POWER GENERATION AND CONSUMPTION

From heating water, to powering lights, to moving our vehicles, nearly every facet of modern human life requires the consumption of power, culminating in nearly 20 TW of power use in recent years. Asia consumes over half of the power derived from combustion of fossil fuels, with Europe and North America each consuming around 20% of the global total. Asia also produces the plurality of power from renewable technologies, such as hydroelectric, wind, and solar, however, North America, South America, and Europe each produce more on a per capita basis. Nuclear energy, however, is primarily produced in Europe, with North America and Asia coming in second and third place, respectively. On a per-capita basis, North America consumes or produces more energy than all other regions considered here, yielding a total power consumption of nearly 10,000 W per person.

**Source:** Energy Information Administration of the United States (2017)
**Notes:** "Renewables" includes hydroelectric, biofuels, biomass (wood), geothermal, wind, and solar. "Fossil fuels" includes coal, oil, and natural gas.

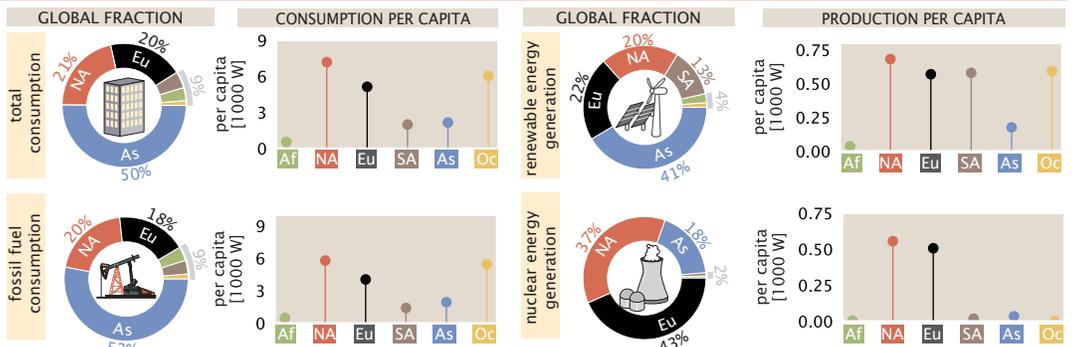

# The Anthropocene by the Numbers — Dimensionless Ratios

Much of our understanding of the scales of things is comparative. When we measure lengths, we do so relative to some intuitive distance that provides context. Our aim here is to present some "yardstick" to measure those numbers presented on pages 1 and 2 in a ratiometric form that compares the magnitude of a given human impact to a natural scale for that same quantity. For example, in considering the use of land by humans, a natural dimensionless way to characterize that number is by comparing it to the total land area of our planet, a comparison that yields what we call the "Terra number."

Similarly, when we consider the entirety of human-made materials, it is natural to compare this mass to the total biomass on our planet. Here we present twelve key human impacts in this dimensionless form. These numbers describe the solid earth, the atmosphere, the biosphere, the oceans, and human resource and energy use, and we hope that our readers will be emboldened to consider their own favorite examples in a similar dimensionless format. Where appropriate, we reference key values using a Human Impacts Database number (HuID) accessible via anthroponumbers.org

### THE TERRA NUMBER

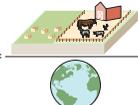

$$Te = \frac{\text{land area used by humans}}{\text{total land area of Earth}} \approx 0.3$$

**The Terra Number** captures the extent to which we have taken control of Earth's terrestrial surface to support our dwellings and, more importantly, our agriculture. Of the $\approx 1.5 \times 10^{14}$ m$^2$ of Earth's surface area that is land, approximately $5 \times 10^{13}$ m$^2$ (HuID: 29582) is used for agriculture, including growing our crops and rearing livestock. Despite being icons of humanity, urban centers occupy between 6.5 and $7.5 \times 10^{11}$ m$^2$ (HuID: 41339, 39341), a total less than 1% of the terrestrial surface. In total, humans directly manage 30% of Earth's terrestrial surface.

### THE DEFORESTATION NUMBER

$$Df = \frac{\text{annual forest loss from human action}}{\text{annual forest loss from wildfire}} \approx 2$$

**The Deforestation Number** reflects the magnitude to which we intentionally clear land of tree cover for production of goods, agriculture, and building our cities relative to the tree cover lost due to wildfire. The intentional clearing of land for production of goods (such as lumber and paper) permanently deforests $\approx 6 \times 10^{10}$ m$^2$ per year. Other intentionally cleared area (which eventually regrows) is comparable in magnitude with $\approx 5 \times 10^{10}$ m$^2$ / yr for forestry (HuID: 38352) and $\approx 7.5 \times 10^{10}$ m$^2$ / yr for shifting agriculture (HuID: 24388). Expansion of urban areas accounts for < 1% of the total annual deforested area, averaging $\approx 2 \times 10^9$ m$^2$ / yr (HuID: 19429). In total, intentional deforestation by humans amounts to $\approx 1.5 \times 10^{11}$ m$^2$ / yr, about twice the area cleared by natural and human-caused wildfires each year ($\approx 7 \times 10^{10}$ m$^2$ / yr, HuID: 92221).

### THE BARNYARD NUMBER

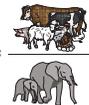

$$By = \frac{\text{mass of terrestrial livestock}}{\text{mass of terrestrial wild animals}} \approx 30$$

**The Barnyard Number** focuses another lens onto the massive agricultural transformation of the planet by comparing the total biomass of terrestrial livestock (e.g. cows, chickens, and pigs) to that of terrestrial wild mammals and birds (e.g. elephants, foxes, and pelicans) [1]. Agricultural intensification of the 20th century has resulted in livestock outweighing all wild terrestrial animals by a factor of $\approx 30$. While poultry make up the vast majority of terrestrial livestock ($\approx 25$ billion individuals, HuID: 94934), they represent a small proportion of livestock biomass. Despite a smaller population of $\approx 1.5$ billion (HuID: 92006), cattle dominate livestock biomass with a total mass of $\approx 1.5 \times 10^{12}$ kg.

### THE NITROGEN NUMBER

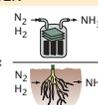

$$N_2 = \frac{\text{mass of } N_2 \text{ fixed via the Haber-Bosch Process}}{\text{mass of } N_2 \text{ fixed biologically}} \approx 1$$

**The Nitrogen Number** illustrates how humans have transformed the global nitrogen cycle to sustain a global population in excess of three billion humans. While molecular nitrogen ($N_2$) is abundant in our atmosphere, nitrogen can only be used by plants in a reactive form such as ammonia ($NH_3$). The 1910 development of the Haber-Bosch process for industrial synthesis of $NH_3$ from $N_2$ was critical for supporting the agricultural needs of a growing human population and for supplying $NH_3$ for chemical and explosive synthesis. Primarily through the Haber-Bosch process, humans synthesize as much reactive nitrogen industrially ($\approx 1.5 \times 10^{11}$ kg / yr, HuID: 61614, 60580) as is synthesized by nitrogen-fixing microbes in terrestrial ecosystems ($\approx 1 \times 10^{11}$ kg per year, HuID: 15205). Beyond influencing the environmental balance of reactive nitrogen, modern synthesis technologies require a sizable amount of energy, contributing significantly to global $CO_2$ emissions.

### THE WATER NUMBER

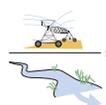

$$Wa = \frac{\text{annual water volume used by humans}}{\text{global annual river discharge volume}} \approx 0.05$$

**The Water Number** captures the magnitude of human water usage relative to global river discharge, a major source of renewable freshwater. Agriculture defines this aspect of human impacts, using $\approx 1.5 \times 10^{12}$ m$^3$ (HuID: 43593) of water annually, accounting for the majority of human water usage. Water used for industrial purposes, including cooling thermoelectric plants amounts to $5.9 \times 10^{11}$ m$^3$ / yr (HuID: 27142), and domestic use is $\approx 6 \times 10^{10}$ m$^3$ / yr (HuID: 69424). In total, annual human water withdrawal is about 5% of global annual river discharge volume, a major source of renewable freshwater. While this is a small fraction, available freshwater is highly variable across the globe and about a third of the human population lives in water stressed areas, where greater than 40% of available freshwater is used.

### THE RIVER NUMBER

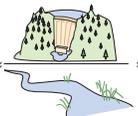

$$Rv = \frac{\text{river volume controlled by humans}}{\text{free-flowing river volume}} \approx 1$$

**The River Number** illustrates the extent to which we have fragmented the free-flowing river systems of the globe for irrigation, flood control, and generation of hydroelectric power. Harnessing this water, however, requires damming the river – thus interrupting its flow and altering the riverine ecosystem. Primarily through damming and construction of channels, humans control $\approx 6 \times 10^{11}$ m$^3$ of water (HuID: 61661), a volume comparable to that freely flowing in unperturbed river systems [2]. Of the free-flowing volume, approximately half is within the Amazon river alone.

### THE EARTH MOVER NUMBER

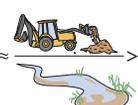

$$Em = \frac{\text{annual mass of earth moved by humans}}{\text{annual mass of earth moved by rivers}} > 15$$

**The Earth Mover Number** describes the mass of sediment moved by humans each year. Through construction, mining, and agriculture, humans move more than $2.5 \times 10^{14}$ kg / yr of sediment a year (HuID: 72899, 59640, 19415, 41496). While there is uncertainty in the total mass of sediment moved through urbanization, the total mass of earth moved by humans is at least 15 times the approximate mass moved each year by the worlds' rivers ($1.3 \times 10^{13}$ kg / yr [3]). This remarkable anthropogenic action rapidly increases erosion rates, leading to increased topsoil loss and turnover, ultimately perturbing natural biogeochemical cycles.

### THE ANTHROPOMASS NUMBER

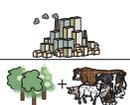

$$An = \frac{\text{total anthropomass}}{\text{total biomass}} \approx 1$$

**The Anthropomass Number** takes stock of our material production by comparing the total quantity of human-made materials to the entirety of the biomass on planet Earth. Around 2020, total human made materials added up to the same mass as the total biomass dry weight ($\approx 1.1 \times 10^{15}$ kg [4]). Concretes and aggregates (such as gravel) dominate the anthropomass, with bricks and asphalt coming in a distant second. Despite their ubiquity, plastics and metals constitute less than 10% of total anthropomass. Altogether, the total amounts to a dizzying $\approx 10^5$ kg of human made mass, or about 20 African bush elephants, per person on the planet.

### THE SOLAR NUMBER

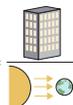

$$Su = \frac{\text{annual human power usage}}{\text{annual incident solar power}} \approx 0.0001$$

**The Solar Number** puts the 20 TW power consumption of human activities (HuID: 31373; 85317) in relief by comparing it to the power incident on our planet from the sun. While humans derive biological energy from food, we derive mechanical and electrical power from various fuel sources like coal, oil, natural gas, and fissile nuclear material. Current global human power usage amounts to an enormous 18 terawatts ($18 \times 10^{12}$ W) and only $\approx 1\%$ comes from solar power (0.12 – 0.20 TW, HuID: 99885; 58303). However, the incident power from the sun dwarfs this number. Of course, it is unlikely that we could ever harness 100% of solar energy incident on our planet, but capturing even 1% of this energy would provide 100 times more power than we currently use.

### THE CO$_2$ NUMBER

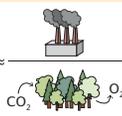

$$CO_2 = \frac{\text{annual mass of anthropogenic } CO_2}{\text{annual mass of naturally removed } CO_2} \approx 2$$

**The CO$_2$ Number** compares the annual amount of human-caused $CO_2$ emissions to the mass of $CO_2$ naturally removed from the atmosphere each year. There are many climate-related consequences of increasing $CO_2$ emissions. Beyond accelerating climate change, $\approx 25\%$ of $CO_2$ released into the atmosphere is absorbed by the oceans, making them appreciably more acidic over time. In recent years, human activities, including burning fossil fuels and making concrete, have led to the release of $\approx 4 \times 10^{13}$ kg of $CO_2$ (HuID: 54608, 24789) into the atmosphere each year. While many natural processes like volcanoes and wildfires release $CO_2$, they are generally accompanied by corresponding sinks that remove even more $CO_2$, like plant photosynthesis. Once all natural processes have been accounted for, a net natural sink of $\approx 2 \times 10^{13}$ kg of $CO_2$ per year remains (HuID: 52670). Thus, the $CO_2$ number quantifies the extent to which human emissions outpace the natural removal of $CO_2$.

### THE METHANE NUMBER

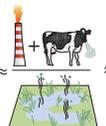

$$Me = \frac{\text{annual mass of anthropogenic } CH_4}{\text{annual mass of natural } CH_4} \approx 1$$

**The CH$_4$ Number** sheds light on the anthropogenic contribution to methane emissions. While $CO_2$ is the most often discussed greenhouse gas, human activities also release substantial amounts of methane ($CH_4$), an even more potent greenhouse gas than $CO_2$. Anthropogenic methane emissions result from fossil fuel extraction, ruminant livestock (mostly cows), rice cultivation, and other sources, totaling $\approx 3$–$4 \times 10^9$ kg per year (HuID: 96837). Natural emissions of CH4, stemming mostly from wetlands and other anaerobic environments, produce a comparable amount ($\approx 2$–$4 \times 10^9$ kg / yr) to anthropogenic emissions (HuID: 56405). Both of these amounts are estimates from models, due to their uncertainty, we report that anthropogenic and natural methane emissions are approximately equal in magnitude.

### THE EXTINCTION NUMBER

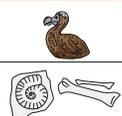

$$Ex = \frac{\text{number of known animal extinctions}}{\text{number of expected natural animal extinctions}} > 10$$

**The Extinction Number** quantifies the deleterious effect that human action has on other species. Over the past 500 years, far more animal species have gone extinct than would be expected due to natural processes. Since 1500, at least 760 animal species have gone extinct (HuID: 44641). Recent estimates of ancient rates of animal extinction predict that tenfold fewer ($\approx 50$) species would have gone extinct over the same period in the absence of humans [5]. It's important to emphasize that these data are incomplete and reflect only a fraction of species that have been assessed for conservation status. **The Extinction Number** therefore represents a lower bound on the degree of modern species loss, with the true value likely being much higher.

# The Anthropocene by the Numbers — Supporting Information

**About:** Here, we present citations and notes corresponding to each quantity assessed here. Each value presented on page 1 is assigned a Human Impacts Database identifier (HuID), accessible via anthroponumbers.org. When possible, primary data sources have been collated and stored as files in comma-separated-value (csv) format on the GitHub repository associated with this snapshot, accessible via DOI: 10.5281/zenodo.4453277 and https://github.com/rpgroup-pboc/human_impacts

## A    SURFACE TEMPERATURE

Surface temperature change relative to 1850–1900 average ≈ 1 – 1.4° C    HuID: 79598, 76539, 12147

**Data Source(s):** HadCRUT.4.6 (Morice et al., 2012, DOI: 10.1029/2011JD017187), GISTEMP v4 (GISTEMP Team, 2020: GISS Surface Temperature Analysis (GISTEMP), version 4. NASA Goddard Institute for Space Studies. Dataset accessed 2020-12-17 at https://data.giss.nasa.gov/gistemp/ & Lenssen et al., 2019, DOI: 10.1029/2018JD029522) and NOAAGlobalTemp v5 (Zhang et al, 2019, DOI: 10.1029/2019EO128229) datasets. **Notes:** The global mean surface temperature captures near-surface air temperature over the planet's land and ocean surface. The value reported represents the spread of the three estimates and their 95% confidence intervals. Temperature changes from all three datasets are expressed relative to the 1850–1900 average temperature from the HadCRUT.4.6 dataset. Since data for the period 1850–1880 are missing in GISTEMP v4 and NOAAGlobalTemp v5, data are centered by setting the 1880–1900 mean of all datasets to the HadCRUT.4.6 mean over the same period.

## B    ANNUAL ICE MELT

glaciers = $(3.0 \pm 1.2) \times 10^{11}$ m³ / yr    HuID: 32459

**Data Source(s):** Intergovernmental Panel on Climate Change (IPCC) 2019 Special Report on the Ocean and Cryosphere in a Changing Climate. Table 2.A.1 on pp. 199–202. **Notes:** Value corresponds to the trend of annual glacial ice volume loss (reported as ice mass loss) from major glacierized regions (2006–2015) based on aggregation of observation methods (original data source: Zemp et al. 2019, DOI:10.1038/s41586-019-1071-0) with satellite gravimetric observations (original data source: Wouters et al. 2019, DOI:10.3389/feart.2019.00096). Ice volume loss was calculated from ice mass loss assuming a standard pure ice density of 920 kg / m³. Uncertainty represents a 95% confidence interval calculated from standard error propagation of the 95% confidence intervals reported in the original sources assuming them to be independent.

ice sheets = $(4.7 \pm 0.4) \times 10^{11}$ m³ / yr    HuID: 95798, 93137

**Data Source(s):** D. N. Wiese et al. 2019 JPL GRACE and GRACE-FO Mascon Ocean, Ice, and Hydrology Equivalent HDR Water Height RL06M CRI Filtered Version 2.0, Ver. 2.0, PO.DAAC, CA, USA. Dataset accessed [2020-Aug-10]. DOI: 10.5067/TEM- SC-3MJ62

**Notes:** Value corresponds to the trends of combined annual ice volume loss (reported as ice mass loss) from the Greenland and Antarctic Ice Sheets (2002–2020) measured by satellite gravimetry. Ice volume loss was calculated from ice mass loss assuming a standard pure ice density of 920 kg / m³. Uncertainty represents one standard deviation and considers only propagation of monthly uncertainties in measurement.

Arctic sea ice = $(3.0 \pm 1.0) \times 10^{11}$ m³ / yr    HuID: 89520

**Data Source(s):** PIOMAS Arctic Sea Ice Volume Reanalysis, Figure 1 of webpage as of October 31, 2020. Original method source: Schweiger et al. 2011, DOI:10.1029/2011JC007084 **Notes:** Value reported corresponds to the trend of annual volume loss from Arctic sea ice (1979–2020). The uncertainty in the trend represents the range in trends calculated from three ice volume determination methods.

## C    SEA ICE EXTENT

extent loss at yearly maximum cover (September) ≈ $8.4 \times 10^{10}$ m² / yr    HuID: 33993

extent loss at yearly minimum cover (March) ≈ $4.0 \times 10^{10}$ m² / yr    HuID: 87741

average annual extent loss = $5.5 \pm 0.2 \times 10^{10}$ m² / yr    HuID: 70818

**Data Source(s):** Comiso et al. 2017, DOI:10.1002/2017JC012768. Fetterer et al. 2017, updated daily. Sea Ice Index, Version 3, Boulder, Colorado USA. NSIDC: National Snow and Ice Data Center, DOI:10.7265/N5K072F8, [Accessed 2020-Oct-19]. **Notes:** Sea ice extent refers to the area of the sea with > 15% ice coverage. Annual value corresponds to the linear trend of annually averaged Arctic sea ice extent from 1979–2015 (Comiso et al 2017) calculated from four different methods. This is in good agreement with the linear trend of annual extent loss calculated by averaging over every month in a given year ($5.5 \times 10^{10}$ m² / yr HuID: 66277). The minimum cover extent loss corresponds to the linear trend of Arctic sea ice extent in September from 1979–2020 and the maximum cover extent loss corresponds to the linear trend of sea ice extent in March from 1979–2020. The Antarctic sea ice extent trend is not shown because a significant long-term trend over the satellite observation period is not observed and short-term trends are not yet identifiable.

## D    MATERIAL PRODUCTION

concrete production ≈ $(2 - 3) \times 10^{13}$ kg / yr    HuID: 25488, 81346, 16995

**Data Source(s):** United States Geological Survey (USGS), Mineral Commodity Summaries 2020, pp. 42–43, DOI:10.3133/mcs2020. Miller et al. 2016, Table 1, DOI:10.1088/1748-9326/11/7/074029. Monteiro et al. 2017, DOI:10.1038/nmat4930. Krausmann et al. 2017, DOI:10.1073/pnas.1613773114 **Notes:** Concrete is formed when aggregate material is bonded together by hydrated cement. The USGS reports the mass of cement produced in 2019 as $4.1 \times 10^{12}$ kg in 2019. As most cement is used to form concrete, cement production can be used to estimate concrete mass using a multiplicative conversion factor of 7 (Monteiro et al.). Miller et al. report that the cement, aggregate and water used in concrete in 2012 sum to $2.3 \times 10^{13}$ kg. Krausmann et al. report an estimated value from 2010 based on a material input, stocks, and outputs model. The value is net annual addition to concrete stocks plus annual waste and recycling to estimate gross production of concrete.

steel production = $(1.4 - 1.9) \times 10^{12}$ kg / yr    HuID: 51453, 44894, 85981

**Data Source(s):** United States Geological Survey (USGS), Mineral Commodity Summaries 2020, pp. 82–83, DOI:10.3133/mcs2020. World Steel Association, World Steel in Figures 2020, p. 6. Krausmann et al. 2017, DOI:10.1073/pnas.1613773114 **Notes:** Crude steel includes stainless steels, carbon steels, and other alloys. The USGS reports the mass of crude steel produced in 2019 as 1900 megatonnes (Mt). The World Steel Association reports a production value of 1869 Mt in 2019. Krausmann et al. report an estimated value from 2010 based on a material input, stocks, and outputs model. The value is net annual addition to steel stocks plus annual waste and recycling to estimate gross production of steel.

plastic production ≈ $4 \times 10^{11}$ kg / yr    HuID: 97241, 25437

**Data Source(s):** Geyer et al. 2017, Table S1, DOI:10.1126/sciadv.1700782. ; Krausmann et al. 2017, DOI:10.1073/pnas.1613773114. **Notes:** Value represents the approximate sum total global production of plastic fibers and plastic resin during the calendar year of 2015. Comprehensive data about global plastic production is sorely lacking. Geyer et al. draw data from various industry groups to estimate total production of different polymers and additives. Some of the underlying data is not publicly available, and data from financially-interested parties is inherently suspect. Krausmann et al. report an estimated value from 2010 based on a material input, stocks, and outputs model. The value is net annual addition to stocks plus annual waste and end-of-life recycling to estimate gross production of plastics.

## E    LIVESTOCK POPULATION

chicken standing population ≈ $2.5 \times 10^{10}$    HuID: 94934

cattle standing population ≈ $1.5 \times 10^{9}$    HuID: 92006

swine standing population ≈ $1 \times 10^{9}$    HuID: 21368

all livestock standing population ≈ $3 \times 10^{10}$    HuID: 43599

**Data Source(s):** Food and Agriculture Organization (FAO) of the United Nations Statistical Database (2020) — Live Animals. **Notes:** Counts correspond to the estimated standing populations in 2018. Values are reported directly by countries. The FAO uses non-governmental statistical sources to address uncertainty and missing (non-reported) data. Reported values are therefore approximations.

## F    SYNTHETIC NITROGEN FIXATION

annual mass of synthetically fixed nitrogen ≈ $1.5 \times 10^{11}$ kg N / yr    HuID: 60580, 61614

**Data Source(s):** United States Geological Survey (USGS), Mineral Commodity Summaries 2020, pp. 116–117, DOI:10.3133/mcs2020. International Fertilizer Association (IFA) Statistical Database (2020) — Ammonia Production & Trade Tables by Region. Smith et al. 2020, DOI: 10.1039/c9ee02873k. **Notes:** Ammonia ($NH_3$) produced globally is compiled by the USGS and IFA from major factories that report output. The USGS estimates the approximate mass of nitrogen in ammonia produced in 2019 as $1.50 \times 10^{11}$ kg N and the International Fertilizer Association reports a production value of $1.50 \times 10^{11}$ kg N in 2019. Nearly all of this mass is produced by the Haber–Bosch process (>96%, Smith et al. 2020). In the United States most of this mass is used for fertilizer, with the remainder being used to synthesize nitrogen-containing chemicals including explosives, plastics, and pharmaceuticals (≈ 88%, USGS Mineral Commodity Summaries 2020).

## G    OCEAN ACIDITY

surface ocean [$H^+$] ≈ 0.2 parts per billion    HuID: 90472

annual change in [$H^+$] = $0.36 \pm 0.03$ %    HuID: 19394

**Data Source(s):** Figures 1–2 of European Environment Agency report CLIM 043 (2020). Original data source of the report is "Global Mean Sea Water pH" from Copernicus Marine Environment Monitoring Service. **Notes:** Reported value is calculated from the global average annual change in pH over years 1985–2018. The average oceanic pH was ≈ 8.057 in 2018 and decreases annually by ≈ 0.002 units, giving a change in [$H^+$] of roughly $10^{-8.056} - 10^{-8.057} \approx 4 \times 10^{-11}$ mol/L or about 0.4% of the global average. [$H^+$] is calculated as $10^{-pH} \approx 10^{-8}$ mol/L or 0.2 parts per billion (ppb) which is calculated by noting that [$H_2O$] ≈ 55 mol / L. Uncertainty for annual change is the standard error of the mean.

## H    LAND USE

agricultural ≈ $5 \times 10^{13}$ m²    HuID: 29582

**Data Source(s):** Food and Agriculture Organization (FAO) of the United Nations Statistical Database (2020) — Land Use. **Notes:** Agricultural land is defined as all land that is under agricultural management including pastures, meadows, permanent crops, temporary crops, land under fallow, and land under agricultural structures (such as barns). Reported value corresponds to 2017 estimates by the FAO.

urban ≈ $(6 - 8) \times 10^{11}$ m²    HuID: 41339, 39341

**Data Source(s):** Florczyk et al. 2019 (https://tinyurl.com/yyxxgtll) and Table 3 of Liu et al. 2018 DOI: 10.1016/j.rse.2018.02.055. **Notes:** Urban land area is determined from satellite imagery. An area is determined to be "urban" if the total population is greater than 5,000 and has a minimum population density of 300 people per km². Reported value gives the range of recent measurements of ≈ $6.5 \times 10^{11}$ m² (2015) and ≈ $(7.5 \pm 1.5) \times 10^{11}$ m² (2010) from Florczyk et al. 2019 and Liu et al. 2018, respectively.

## I    RIVER FRAGMENTATION

global fragmented river volume ≈ $6 \times 10^{11}$ m³    HuID: 61661

**Data Source(s):** Grill et al. 2019 DOI: 10.1038/s41586-019-1111-9. **Notes:** Value corresponds to the water volume contained in rivers that fall below the connectivity threshold required to classify them as free-flowing. Value considers only rivers with upstream catchment areas greater than 10 km² or discharge volumes greater than 0.1 m³ per second. The ratio of global river volume in disrupted rivers to free-flowing rivers is approximately 0.9. The exact value depends on the cutoff used to define a "free-flowing" river. We direct the reader to the source for thorough detail.

## J    HUMAN POPULATION

urban-dwelling fraction of population ≈ 55%    HuID: 93995

total population ≈ $7.6 \times 10^{9}$    HuID: 85255

**Data Source(s):** Food and Agricultural Organization (FAO) of the United Nations Report on Annual Population, 2019. **Notes:** Value for total population in 2018 comes from a combination of direct population reports from country governments as well as inferences of underreported or missing data. The definition of "urban" differs between countries and the data does not distinguish between urban and suburban populations despite substantive differences between these land uses (Jones and Kammen 2013, doi: 10.1021/es4034364). As explained by the United Nations population division, "When the definition used in the latest census was not the same as in previous censuses, the data were adjusted whenever possible so as to maintain consistency." Rural population is computed from this fraction along with the total human population, implying that the total population is composed only of "urban" and "rural" communities.

## K    GREENHOUSE GAS EMISSIONS

anthropogenic $CO_2$ = $(4.25 \pm 0.33) \times 10^{13}$ kg $CO_2$ / yr    HuID: 24789, 54608, 98043, 60670

**Data Source(s):** Table 6 of Friedlingstein et al. 2019, DOI: 10.5194/essd-11-1783-2019. Original data sources relevant to this study compiled in Friedlingstein et al.: 1) Gilfillan et al. https://energy.appstate.edu/CDIAC 2) Average of two bookkeeping models: Houghton and Nassikas 2017 DOI: 10.1002/2016GB005546; Hansis et al. 2015 DOI: 10.1002/2014GB004997) Dlugokencky and Tans, NOAA/GML https://www.esrl.noaa.gov/g-md/ccgg/trends/. **Notes:** Value corresponds to total $CO_2$ emissions from fossil fuel combustion, industry (predominantly cement production), and land-use change during calendar year 2018. Emissions from land-use change are due to the burning or degradation of plant biomass. In 2018, $1.88 \times 10^{13}$ kg $CO_2$ / yr accumulated in the atmosphere, reflecting the balance of emissions and $CO_2$ uptake by plants and oceans. Uncertainty corresponds to one standard deviation.

# The Anthropocene by the Numbers — Supporting Information

## K — GREENHOUSE GAS EMISSIONS (CONTINUED)

anthropogenic $CH_4$ = (3.4 – 3.9) × $10^{11}$ kg $CH_4$ / yr — HuID: 96837, 30725

Data Source(s): Table 3 of Saunois, et al. 2020. DOI: 10.5194/essd-12-1561-2020. Notes: Value corresponds to 2008–2017 decadal average mass of $CH_4$ emissions from anthropogenic sources. Includes emissions from agriculture and landfill, fossil fuels, and burning of biomass and biofuels, but other inventories of anthropogenic methane emissions are also considered. Reported range represents the minimum and maximum estimated emissions from a combination of "bottom–up" and "top–down" models.

anthropogenic $N_2O$ = 1.1 (+0.6, – 0.5) × $10^{10}$ kg $N_2O$ / yr — HuID: 44575

Data Source(s): Table 1 of Tian, H., et al. 2020. DOI: 10.1038/s41586-020-2780-0. Notes: Value corresponds to annualized $N_2O$ emissions from anthropogenic sources in the years 2007–2016. The value reported in the source is 7.3 (4.2, 11.4) Tg N / year. This is converted to a mass of $N_2O$ using the fact that N ≈ 14/22 of the mass of $N_2O$. Reported value is mean with the uncertainty bounds (+,–) representing the maximum and minimum values observed in the 2007–2016 time period.

## L — WATER WITHDRAWAL

agricultural withdrawal = 1.3 × $10^{12}$ $m^3$ / yr — HuID: 84545, 43593, 95345
industrial withdrawal = 5.9 × $10^{11}$ $m^3$ / yr — HuID: 27142
domestic withdrawal = 5.4 × $10^{10}$ $m^3$ / yr — HuID: 69424
total withdrawal = (1.7 – 2.2) × $10^{12}$ $m^3$ / yr — HuID: 27342, 68004

Data Source(s): Figure 1 of Qin et al. 2019. DOI: 10.1038/s41893-019-0294-2. AQUASTAT Main Database, Food and Agriculture Organization of the United Nations Notes: "Agricultural" and "total" withdrawal include one value from Qin et al. (who reports "consumption") and one value from the AQUASTAT database. Industrial water withdrawal is from AQUASTAT and domestic withdrawal value is from Qin et al. Values in AQUASTAT are self-reported by countries and have missing values from some countries, probably accounting for a few percent underreporting. All values represent withdrawals. For agricultural and domestic, water withdrawal is assumed to be the same as water consumption as reported in Qin et al.

## M — SEA LEVEL RISE

added water = 1.97 (+0.36, –0.34) mm / yr — HuID: 97108
thermal expansion = 1.19 (+0.25, –0.24) mm / yr — HuID: 97688
total observed sea-level rise = 3.35 (+0.47, –0.44) mm / yr — HuID: 81373

Data Source(s): Table 1 of Frederikse et al. 2020. DOI:10.1038/s41586-020-2591-3. Notes: Values correspond to the average global sea level rise for the years 1993 – 2018. "Added water" (barystatic) change includes effects from meltwater from glaciers and ice sheets, added mass from sea-ice discharge, and changes in the amount of terrestrial water storage. Thermal expansion accounts for the volume change of water with increasing temperature. Values for "thermal expansion" and "added water" come from direct observations of ocean temperature and gravimetry/altimetry, respectively. Total sea level rise is the observed value using a combination of measurement methods. "Other sources" reported on page 1 accounts for observed residual sea level rise not attributed to a source in the model. Values in brackets correspond to the upper and lower bounds of the 90% confidence interval.

## N — TOTAL POWER USE

global power use ≈ 19 – 20 TW — HuID: 31373, 85317

Data Source(s): bp Statistical Review of World Energy, 2020; U.S. Energy Information Administration, 2020. Notes: Value represents the sum of total primary energy consumed from oil, natural gas, coal, and nuclear energy and electricity generated by hydroelectric and other renewables. Value is calculated using annual primary energy consumption as reported in data sources assuming uniform use throughout a year, yielding ≈ 19 – 20 TW.

## O — TREE COVERAGE AREA LOSS

commodity-driven deforestation = (5.7 ± 1.1) × $10^{10}$ $m^2$ / yr — HuID: 96098
forestry = (5.4 ± 0.8) × $10^{10}$ $m^2$ / yr — HuID: 38352
urbanization = (2 ± 1) × $10^9$ $m^2$ / yr — HuID: 19429
shifting agriculture = (7.5 ± 0.9) × $10^{10}$ $m^2$ / yr — HuID: 24388
wildfire = (7.2 ± 1.3) × $10^{10}$ $m^2$ / yr — HuID: 92221
total loss ≈ 2 × $10^{11}$ $m^2$ / yr — HuID: 78576

Data Source(s): Table 1 of Curtis et al. 2018 DOI:10.1126/science.aau3445. Hansen et al. 2013 DOI:10.1126/science.1244693. Global Forest Watch, 2020. Reported values in source correspond to total loss from 2001 – 2015. Values given are averages over this 15 year window. Notes: Commodity-driven deforestation is "long-term, permanent, conversion of forest and shrubland to a non-forest land use such as agriculture, mining, or energy infrastructure." Forestry is defined as large-scale operations occurring within managed forests and tree plantations with evidence of forest regrowth in subsequent years. Urbanization converts forest and shrubland for the expansion and intensification of existing urban centers. Disruption due to "shifting agriculture" is defined as "small- to medium-scale forest and shrubland conversion for agriculture that is later abandoned and followed by subsequent forest regrowth". Disruption due to wildfire is "large-scale forest loss resulting from the burning of forest vegetation with no visible human conversion or agricultural activity afterward". Uncertainty corresponds to the 95% confidence interval. Uncertainty is approximate for "urbanization" as the source reports an ambiguous error of "± <1%".

## P — POWER FROM FOSSIL FUELS

natural gas = 4.5 – 4.8 TW — HuID: 49947, 86175
oil = 6.1 – 6.6 TW — HuID: 4121, 39756
coal = 5.0 – 5.5 TW — HuID: 10400, 60490
total = 16 – 17 TW — HuID: 29470, 29109

Data Source(s): bp Statistical Review of World Energy, 2020. U.S. Energy Information Administration, 2020. Notes: Values are self-reported by countries. values from bp Statistical Review correspond to 2019 whereas values from the EIA correspond to 2018 estimates. Reported TW are computed from primary energy (e.g. kg coal) units assuming uniform use throughout the year. Oil volume includes crude oil, shale oil, oil sands, condensates, and natural gas liquids separate from specific natural gas mining. Natural gas value excludes gas flared or recycled and includes natural gas produced for gas-to-liquids transformation. Coal value includes 2019 value exclusively for solid commercial fuels such as bituminous coal and anthracite, lignite and subbituminous coal, and other solid fuels. This includes coal used directly in power production as well as coal used in coal-to-liquids and coal-to-gas transformations.

## Q — POWER FROM RENEWABLE RESOURCES

wind ≈ 0.36–0.39 TW — HuID: 30581, 85919
solar ≈ 0.18 – 0.20 TW — HuID: 99885, 58303
hydroelectric = 1.2 TW — HuID: 15765, 50558
total renewable power ≈ 1.9 – 2.1 TW — HuID: 75741, 20246

Data Source(s): bp Statistical Review of World Energy, 2020. U.S. Energy Information Administration, 2020. Notes: Reported values correspond to estimates for the 2019 and 2018 calendar years for bp and EIA sources, repsectively. Renewable resources are defined as wind, geothermal, solar, biomass and waste. Hydroelectric, while presented here, is not defined as a renewable in the bp dataset. All values are reported as input-equivalent energy, meaning the input energy that would have been required if the power was produced by fossil fuels. BP reports that fossil fuel efficiency used to make this conversion was ≈ 40% in 2017.

## R — FOSSIL FUEL EXTRACTION

volume of natural gas = (3.9 – 4.0) × $10^{12}$ $m^3$ / yr — HuID: 11468, 20532
volume of oil = (5.5 – 5.8) × $10^9$ $m^3$ / yr — HuID: 66789, 97719
mass of coal = (7.8 – 8.1) × $10^{12}$ kg / yr — HuID: 78435, 48928

Data Source(s): bp Statistical Review of World Energy, 2020. U.S. Energy Information Administration, 2020. Notes: Oil volume includes crude oil, shale oil, oil sands, condensates, and natural gas liquids separate from specific natural gas mining. Natural gas value excludes gas flared or recycled and includes natural gas produced for gas-to-liquids transformation. Coal value includes solid commercial fuels such as bituminous coal, anthracite, lignite, subbituminous coal, and other solid fuels. All values from bp Statistical Review correspond to 2019 whereas values from the EIA correspond to 2018 estimates.

## S — OCEAN WARMING

heat uptake by ocean ≈ 346 ± 51 TW — HuID: 94108
upper ocean (0 – 700 m) temperature increase since 1960 = 0.18 – 0.2 °C — HuID: 69674, 72086

Data Source(s): Table S1 of Cheng et al. 2017. doi: 10.1126/sciadv.1601545. NOAA National Centers for Environmental Information, 2020. doi:10.1029/2012GL051106. Notes: Heat uptake reported is the average over time period 1992–2015 with 95% confidence intervals. Range of temperatures reported captures the 95% confidence interval of temperature increase for the period 2015–2019 with respect to the 1958–1962 mean. Temperature change is considered in the upper 700 m because sea surface temperatures have high decadal variability and are a poor indicator of ocean warming; see Roemmich et al. 2015, doi: 10.1038/NCLIMATE2513.

## T — POWER FROM NUCLEAR FISSION

nuclear power ≈ 0.79–0.89 TW — HuID: 48387

Data Source(s): bp Statistical Review of World Energy, 2020. U.S. Energy Information Administration, 2020. Notes: Values are self-reported by countries and correspond to estimates for 2019 and 2018 calendar year for bp and EIA data, respectively. Values are reported as 'input-equivalent' energy, meaning the energy needed to produce a given amount of power if the input were a fossil fuel, which is converted to TW here. This is calculated by multiplying the given power by a conversion factor representing the efficiency of power production by fossil fuels. In 2017, this factor was about 40%.

## U — NUCLEAR FALLOUT

anthropogenic $^{239}Pu$ and $^{240}Pu$ from weapons testing ≈ 1.4 × $10^{11}$ kg / yr — HuID: 42526

Data Source(s): Table 1 in Hancock et al. 2014 doi: 10.1144/SP395.15. Fallout in activity from UNSCEAR 2000 Report on Sources and Effects of Ionizing Radiation Report to the UN General Assembly –- Volume 1. Notes: The approximate mass of Plutonium isotopes $^{239}Pu$ and $^{240}Pu$ released into the atmosphere from the ≈ 500 above-ground nuclear weapons tests conducted between 1945 and 1980. Naturally occurring $^{239}Pu$ and $^{240}Pu$ are rare, meaning that nearly all contemporary labile plutonium comes from human activity. (Taylor 2001,doi: 10.1016/S1569-4860(01)80003-6) The total mass of radionuclides released is ≈ 3300 kg with a combined radioactive fallout of ≈11 PBq. These values do not represent the entire $^{239+240}Pu$ globally distributed mass as it excludes non-weapons sources.

## V — CONTEMPORARY EXTINCTION

animal species extinct since 1500 > 750 — HuID: 44641
plant species extinct since 1500 > 120 — HuID: 86866

Data Source(s): The IUCN Red List of Threatened Species. Version 2020-2. Notes: Values correspond to absolute lower-bound count of animal extinctions caused over the past ≈ 520 years. Of the predicted ≈ 8 million animal species, the IUCN databases catalogues only ≈ 900,000 with only ≈ 75,000 being assigned a conservation status. Representation of plants and fungi is even more sparse with only ≈40,000 and ≈285 being assigned a conservation status, respectively. The number of extinct animal species is undoubtedly higher than these reported values, as signified by an inequality symbol (>).

## W — EARTH MOVING

waste and overburden from coal mining ≈ 6.5 × $10^{13}$ kg / yr — HuID: 72899
earth moved from urbanization > 1.4 × $10^{14}$ kg / yr — HuID: 59640

Data Source(s): Supplementary table 1 of Cooper et al. 2018. DOI: doi.org/gfwfhd. Notes: Coal mining waste and overburden mass is calculated given commodity-level stripping ratios (mass of overburden/waste per mass of coal resource mined) and reported values of global coal production by type. Urbanization mass is presented as a lower bound estimate of the mass of earth moved from global construction projects. This comes from a conservative estimate that the ratio of the mass of earth moved per mass of cement/concrete used in construction globally is 2:1. This value is highly context dependent and we encourage the reader to read the source material for a more thorough description of this estimation.

erosion from agricultural land > 1.2 – 2.4 × $10^{13}$ kg / yr — HuID: 19415, 41496

Data Source(s): Pg. 377 of Wang and Van Oost 2019. DOI: 10.1177/0959683618816499. Pg. 21996 of Borrelli et al. 2020 DOI: 10.1073/pnas.2001403117. Notes: Cumulative sediment mass loss over history of human agriculture due to accelerated erosion is estimated to be ≈ 30,000 Gt. Recent years have an estimated erosion rate ranging from 12 Pg / yr (Wang and Van Oost) to ≈ 24 Pg / yr (Borrelli et al.). Values come from computational models conditioned on time-resolved measurements of sediment deposition in catchment basins.

We are incredibly grateful for the generosity of a wide array of experts for their advice, suggestions, and criticism of this work. Specifically, we thank Suzy Beeler, Lars Bildsten, Justin Bois, Chris Bowler, Matthew Burgess, Ken Caldeira, Jörn Callies, Sean B. Carroll, Ibrahim Cissé, Joel Cohen, Michelle Dan, Bethany Ehlmann, Gidon Eshel, Paul Falkowski, Daniel Fisher, Thomas Frederikse, Neil Fromer, Eric Galbraith, Lea Goentoro, Evan Groover, John Grotzinger, Soichi Hirokawa, Greg Huber, Christina Hueschen, Bob Jaffe, Elizabeth Kolbert, Thomas Lecuit, Raphael Magarik, Jeff Marlow, Brad Marston, Jitu Mayor, Elliot Meyerowitz, Lisa Miller, Dianne Newman, Luke Oltrogge, Nigel Orme, Victoria Orphan, Marco Pasti, Pietro Perona, Noam Prywes, Stephen Quake, Hamza Raniwala, Manuel Razo-Mejia, Thomas Rosenbaum, Benjamin Rubin, Alex Rubinsteyn, Shyam Saladi, Tapio Schneider, Murali Sharma, Alon Shepon, Arthur Smith, Matthieu Talpe, Wati Taylor, Julie Theriot, Tadashi Tokieda, Cat Triandafillou, Sabah Ul-Hasan, Tine Valencic, and Ned Wingreen. We also thank Yue Qin for sharing data related to global water consumption. Many of the topics in this work began during the Applied Physics 150C course taught at Caltech during the early days of the COVID-19 pandemic. This work was supported by the Resnick Sustainability Institute at Caltech and the Schwartz-Reisman Collaborative Science Program at the Weizmann Institute of Science.